\documentclass{aastex}
\usepackage[onecolumn]{emulateapj5}
\usepackage{epsfig}
\usepackage{apjfonts}

\submitted{ApJ in Press; Received: 2000 September 15; Accepted 2000 October 23}

\shorttitle{Blazar Proper Motions}
\shortauthors{Homan et al.}

\begin{document}

\title{Parsec-Scale Blazar Monitoring: Proper Motions}

\author{Daniel C. Homan\altaffilmark{1}, Roopesh Ojha\altaffilmark{2}, 
John F. C. Wardle\altaffilmark{1}, and David H. Roberts\altaffilmark{1},}
\affil{Physics Department MS057, Brandeis University}
\affil{Waltham, MA 02454} 
\author{Margo F. Aller\altaffilmark{3}, Hugh D. Aller\altaffilmark{3}, and 
Philip A. Hughes\altaffilmark{3}}
\affil{Radio Astronomy Observatory, University of Michigan}
\affil{Ann Arbor, MI 48109} 
\altaffiltext{1}{dhoman@brandeis.edu,jfcw@quasar.astro.brandeis.edu,roberts@brandeis.edu}
\altaffiltext{2}{Harvard Smithsonian Center
                 for Astrophysics; 60 Garden St. MS 78; Cambridge MA 02138; 
                 rojha@cfa.harvard.edu}
\altaffiltext{3}{margo,hugh,hughes@astro.lsa.umich.edu}

\begin{abstract}
We present proper motions obtained from a dual frequency, 
six-epoch, VLBA polarization experiment monitoring a sample 
of 12 blazars. The observations were made at 15 GHz and 22 GHz at 
bi-monthly intervals over 1996. Ten of the 
eleven sources for which proper motion could be reliably 
determined are superluminal.  Only J2005+77 has no superluminal 
components. Three sources (OJ\,287, J1224+21, and
J1512$-$09) show motion faster than $10h^{-1}$c, requiring 
$\gamma_{pattern}$ of at least $10h^{-1}$ 
($H_0 = 100h$ km s$^{-1}$ Mpc$^{-1}$). We compare our results
to those in the literature and find motions outside the previously
observed range for four sources.  While some jet components exhibit 
significant non-radial motion, most motion is radial.  In at 
least two sources there are components moving radially at significantly
different structural position angles. In five of 
six sources (3C\,120, J1224+21, 3C\,273, 3C\,279, J1512$-$09, and J1927+73) 
that have multiple components with measurable proper motion, 
the innermost component is significantly slower than the others,
suggesting that acceleration occurs in the jet. 
In the motions of individual components we observe at least 
one decelerating motion and two 
``bending'' accelerations which tend to align their motions 
with larger scale structure.  
We also discuss in detail our procedures for 
obtaining robust kinematical results from multi-frequency VLBI 
data spanning several epochs. 
\end{abstract}


\section{Introduction}
\label{s:intro}

Apparent superluminal motion is one of the key results to emerge
from Very Long Baseline Interferometry (VLBI). Since its initial 
detection in 1971 \citep{W71,CCPS71}, the number of compact radio sources 
where components appear to move apart at a transverse velocity that 
exceeds the speed of light has been increasing rapidly \citep{VC94}. 
Insofar as they confirm relativistic jet speeds, measurements of 
superluminal motion are integral to our current explanations of the 
high energies inferred for quasars. 

The most widely accepted explanation of superluminal motion 
\citep{R66,BK79} postulates a collimated pair of jets of plasma 
expanding from the nuclear region of an active galactic nucleus 
(AGN). If one of these jets is pointed close to our line of sight, 
contraction of the apparent time scale creates the illusion of 
superluminal transverse motions. For a jet oriented at an angle $\Theta$
relative to the observer, a pattern moving at $\beta_p$ down the jet will
appear to be moving at a speed 
$\beta_{app} = \beta_p\sin\Theta/(1-\beta_p\cos\Theta)$, which can
greatly exceed unity, reaching a maximal value of $\gamma_p = (1-\beta_p^2)^{-1/2}$
when $\beta_p = \cos\Theta$. 
``Doppler favoritism,'' the boost in the flux from such a closely 
aligned jet, makes superluminal sources some of the brightest 
in the radio sky. 

Even as the number of known superluminal sources has been rising,
many of the basic questions about them remain open. This is in large 
part due to a number of difficulties in measuring proper motions with 
precision.  Experiments with {\em ad hoc} arrays can be difficult to 
set up and are difficult to repeat at regular and frequent intervals. 
They are usually made at only one frequency with no polarization 
information. They often have sparse (u,v)-coverage and use different 
sets of dissimilar antennas over different epochs. With the exception 
of a few, well observed sources, most of the known proper motions are 
subject to at least some of these problems.

All the issues listed above impact on the fundamental problem in 
proper motion study - the identification of components (i.e., coherent 
source structure) between epochs.  Frequent monitoring plays a key role, 
for example
in preventing multiple, quickly moving components from being mis-identified
as a single slowly moving component (sometimes called strobing).  By 
providing independent observations,
multiple frequencies are very helpful in resolving confusing behavior and 
adding confidence to component identifications.  Polarization
information provides additional identifying characteristics to 
components and can be vital when the source behavior is particularly 
complex (see OJ\,287 below, for example).  Finally, having consistent (u,v)-coverage 
from epoch to epoch is important to the consistent calibration and modeling 
of source structure.

Here we present proper motion results from six epochs of VLBI 
observation of 12 blazars. These sources were selected from
among the most variable sources being monitored by the University of
Michigan Radio Astronomy Observatory (UMRAO).  They were
chosen to study the rapid evolution of parsec-scale total intensity and
polarization structure in the most active blazars.  
The observations were made at two month intervals
with NRAO's (National Radio Astronomy Observatory) Very Long Baseline 
Array (VLBA)\footnote{The National Radio 
Astronomy Observatory is a facility of the National Science Foundation
operated under cooperative agreement by Associated Universities, Inc.} at 
15 and 22 GHz with full polarization information.  
The VLBA \citep{N95,T95} is a group of 
10 identical telescopes, with identical back-ends, that are  
located across the United States so as to optimize (u,v)-coverage.  
Results from other aspects 
of this monitoring program have been presented elsewhere (e.g. \citet{WHOR98,
HW99, HW00}) and others are in preparation. In particular, the entire dataset in the
form of images and tabular data will be presented in Ojha et al. (in prep.). 
We will also explore the total intensity and polarization structure and variability of
these sources in a third paper (Ojha et al., in prep.).  A fourth paper 
(Aller et al., in prep.) will examine the relationship between flux 
and polarization outbursts and component origin and evolution.

Key questions we investigate here include: (1) How do our observed proper motions
compare to those obtained from quasi-annual observations, often made at lower
frequencies (e.g., $5$ GHz)? (2) Are there significant non-radial motions
of jet components? (3) How do the speeds of different components in the
same jet compare?  Is there a systematic dependence of velocity 
on position in the jet? 
(4) Are there accelerations in the motions of individual jet components, 
either along the direction of motion or perpendicular to it?  

Section \S{\ref{s:obs}} describes our sample, data reduction, and 
model-fitting procedures.  Conventions used throughout the paper are 
detailed in \S{\ref{s:conv}}. Proper motion results on individual
sources are presented in \S{\ref{s:results}}. In \S{\ref{s:discuss}} 
we explore the above questions in the context of our 
sample as a whole, and our conclusions appear in \S{\ref{s:conclude}}.

%
%
%

\section{Observations}
\label{s:obs}

\subsection{The Sample}
We used the VLBA to conduct a series of 6 experiments, each of 24
hour duration, at (close to) two month intervals during the year
1996. The observations were made at 15 GHz ($\lambda$2 cm, U-band) 
and 22 GHz ($\lambda$1.3 cm, K-band).
We observed 11 target sources for six epochs and one (J1224+21) for 
only the last five epochs.  These sources are listed in table \ref{t:Sources}.
The epochs, labeled ``A'' through ``F'' throughout this paper, are listed 
in table \ref{t:Epochs}. For four sources we have additional observations 
at a later date, 1997.94, that we refer to as epoch ``G''. 

%


\begin{table*}[t]
\begin{scriptsize}
\begin{center}
\tablenum{1}
\caption[]{\label{t:Sources}Source Information\\}
\begin{tabular}{cccccccccc}
\tableline\tableline
J2000.0 & J1950.0 & Other Names & Redshift & Classification &\\
\tableline 
J0433+053 & B0430+052 & 3C\,120, II Zw 14 & 0.033 & Sy 1\\
J0530+135\tablenotemark{b} & B0528+134 & PKS 0528+134 & 2.060 & Quasar\\
J0738+177 & B0735+178 & OI\,158, DA\,237, PKS 0735+178 & 0.424 \tablenotemark{a} & BL\\
J0854+201\tablenotemark{b} & B0851+202 & OJ\,287 & 0.306 & BL\\
J1224+212\tablenotemark{b} & B1222+216 & 4C\,21.35 & 0.435 & Quasar\\
J1229+020 & B1226+023 & 3C\,273 & 0.158 & Quasar\\
J1256$-$057\tablenotemark{b} & B1253$-$055 & 3C\,279 & 0.536 & Quasar\\
J1310+323 & B1308+326 & OP\,313 & 0.996 & Quasar/BL\\
J1512$-$090 & B1510$-$089 & OR\,$-$017 & 0.360 & Quasar\\
J1751+09  & B1749+096 & OT\,081, 4C\,09.56 & 0.322 & BL\\
J1927+739 & B1928+738 & 4C\,73.18 & 0.302 & Quasar\\
J2005+778 & B2007+777 &  & 0.342 & BL\\
\tableline 
\end{tabular} 

\tablenotetext{a}{Lower limit}
\tablenotetext{b}{Also observed at epoch 1997.94 (G)}
\end{center}
\end{scriptsize} 
\end{table*} 


\begin{table*}[t]
\begin{scriptsize}
\begin{center}
\tablenum{2}
\caption[]{\label{t:Epochs}Epochs of Observation\\}
\begin{tabular}{cccccccccc}
\tableline \tableline 
Epoch & Day & Label & Notes \\
\tableline 
1996.05 & 19 Jan & A &	\tablenotemark{a} \\
1996.23 & 22 Mar & B &	\tablenotemark{b} \\
1996.41 & 27 May & C &	\tablenotemark{c} \\
1996.57 & 27 Jul & D &                    \\
1996.74 & 27 Sep & E &	\tablenotemark{d}$\quad$\tablenotemark{e} \\
1996.93 & 06 Dec & F &  \tablenotemark{f}                  \\
1997.94 & 07 Dec & G &  \tablenotemark{g}                  \\
\tableline
\end{tabular} 

\tablenotetext{a}{North Liberty antenna off-line for entire experiment}
\tablenotetext{b}{Owens Valley antenna off-line for entire experiment}
\tablenotetext{c}{No fringes found to the Kitt Peak antenna}
\tablenotetext{d}{North Liberty antenna off-line for second half of experiment}
\tablenotetext{e}{Some data loss from the Owens Valley antenna}
\tablenotetext{f}{Numerous problems spread over several antennas; poor
data quality compared to the other epochs.}
\tablenotetext{g}{For four sources only.}
\end{center}
\end{scriptsize} 
\end{table*}

The sources were chosen from those regularly monitored by the
University of Michigan Radio Astronomy Observatory (UMRAO) in total
intensity and polarization at $4.8$, $8.0$, and $14.5$ GHz.  They were
selected according to the following criteria. (1) High 
total intensity: The weakest sources are about 1 Jy, the most powerful
as much as 22 Jy. (2) High polarized flux: Typically over 50 mJy.
(3) Violently variable: In both total and polarized intensity. Such 
sources are likely to be under-sampled by annual VLBI. 
(4) Well distributed in right ascension: This 
allowed us to make a optimal observing schedule.   
 
About 112 of the UMRAO sources meet the first three of the above
criterion. The 12 actually selected were the strongest, most violently 
variable sources, subject to the fourth criteria. Clearly these sources
do not comprise a ``complete sample'' in any sense.  

\subsection{Data Calibration} 
\label{s:reduce}

The frequency agility and high slew speeds of the VLBA antennas were used to 
schedule our observations to generate maximal (u,v)-coverage. Scan lengths were
kept short (13 minutes for the first two epochs and 5.5 minutes for the last
four or five), with a switch in frequency at the end of each scan. In addition, 
scans of neighboring sources were heavily interleaved at the cost of some additional 
slew time.  Each source was observed for approximately 45 minutes per 
frequency at each epoch.  

The data were correlated on the VLBA correlator in
Socorro, NM.  After correlation, the data were distributed on DAT tape to 
Brandeis University where they were loaded into NRAO's Astronomical Imaging 
Processing System (AIPS) \citep{BG94,G88} and calibrated using standard 
techniques for VLBI polarization observations, e.g., \citep{C93,RWB94}.  
For a detailed description of our calibration steps see Ojha et al. (in prep.).

\subsection{Modeling the data}
\label{s:model}

The final CLEAN images of our sources present a wealth of information.  In
many ways, the images contain too much information to be simply parameterized
for quantitative analysis.  To study proper motions, we used the model fitting 
capabilities of the DIFMAP software package \citep{SPT94,SPT95} to fit 
the sources with a number of discrete Gaussian components. The fitting was 
done directly on the final, self-calibrated visibility data 
(i.e., in the (u,v)-plane).  
Obtaining a discrete list of component properties allows for robust mathematical 
analysis of proper motions, one of the key goals of our observing program.

Operationally, components are simply two dimensional Gaussian fits to some or 
all of the visibility data from a source.
We are agnostic about the physical significance of components.  
Components which correspond to compact, enhanced regions of
brightness in the jet are the easiest to follow for proper motion
analysis.  They could be shock induced dense regions traveling along the 
jet, they could
result from variation in the Doppler factor as the jet bends, or slight
variations in speeds could lead to faster plasma catching up with slower
plasma, increasing brightness. 
 
Our approach to model fitting was empirical and conservative. We chose to 
fit the visibilities instead of modeling in the image plane which 
was one step removed from the data. 
We fit the visibility data with elliptical Gaussians (though point 
sources were used occasionally) as this made 
the fewest assumptions about the nature of the components. We sought to obtain 
the simplest possible model, i.e., the model with the least number of 
components
that gave a good fit to the data as judged by a relative chi-squared
statistic. In addition, for a fit to a source to be considered acceptable we 
required the components to contain $\gtrsim$ 95\% of the total flux 
and a convolution of the model components with the beam to be similar to the 
CLEAN image of the source.  In some cases it was possible to fit
an additional component, but we have not done so unless its presence
led to a significant improvement in the quality
of the fit. Several techniques were used to ensure that a fit was 
not merely a local minimum; these included trying different starting points and 
deliberately perturbing the final model.  

No attempt was made to ``drive'' the fit towards previous models where
such models exist in the literature. Indeed, we intentionally remained ignorant
of such models to keep our work unbiased during this part of the analysis. We did
try to maintain consistency between the model-fits, across both epoch and 
frequency of our observations; however, the primary goal was always 
to obtain the best representation of the data in that epoch and at 
that frequency.

Modeling a jet with Gaussian components will work best with
sources that are dominated by discrete, well separated structures.
We had the most difficulty in fitting sources that have complex
morphology with a large fraction of flux in diffuse, extended structures
(e.g., 3C\,120, described below).
The relative flux densities, positions, and dimensions of the Gaussian
components that make up a fit can be strongly correlated, particularly
when jet features are closely spaced or poorly defined.  
The chief problems that such ``cross-talk'' between components lead to are
(1) poor modeling of weak structures, e.g., the northern bar in J0738+17,
and (2) poor modeling of structures that are close  
to each other.  If one of these is very bright, a dimmer companion may
have its position seriously skewed, e.g., component U2 (K2) in 3C\,279. 

Given these issues, it is clear that obtaining a good fit to a source is 
only the first step; deciding which components are reliable tracers of the 
motion of jet structures is critical.  We consider a component to be 
kinematically useful if it meets the following criteria: 

\begin{itemize}

\item It is consistent at our two frequencies: The difference in resolution,
sensitivity and, occasionally, (u,v)-coverage (due to failure of a receiver at one
frequency) often give useful perspectives on the ``reality'' of jet features.
Occasionally we follow a feature well at one frequency and not the other; these
cases are described in the text and marked in the tables. 

\item It is consistent over many epochs, either steady in shape and flux or
changing smoothly. We often fit components in a particular part of a jet at only a
few of our epochs. While such transitory components represent real flux in the
jet, in this paper we generally do not use such components to derive kinematic
information.

\item The position of the component cannot be strongly biased by a close 
neighbor. We have found that position and flux can often be exchanged in the
model-fitting of two closely spaced components. A weak neighbor may have
its properties distorted by a strong component, particularly if
the one (or both) components has a large angular extent (e.g., the relationship of U2 (K2)
and U1 (K1) in 3C\,279).

\end{itemize}

Here we use 3C\,120 to illustrate the  above issues in a particularly difficult
case. In our sample, this was the most difficult source to model-fit robustly
because its structure did not lend itself to simple representation as a collection
of well-separated, discrete components. The intensity near the core declines in a
smooth manner, and there is appreciable underlying emission all along
the jet. While there are edges and lumps superimposed on this smooth background,
we were able to fit such structures reliably only when they were very prominent. 
 
\begin{figure*}
\epsscale{0.5}
\plotone{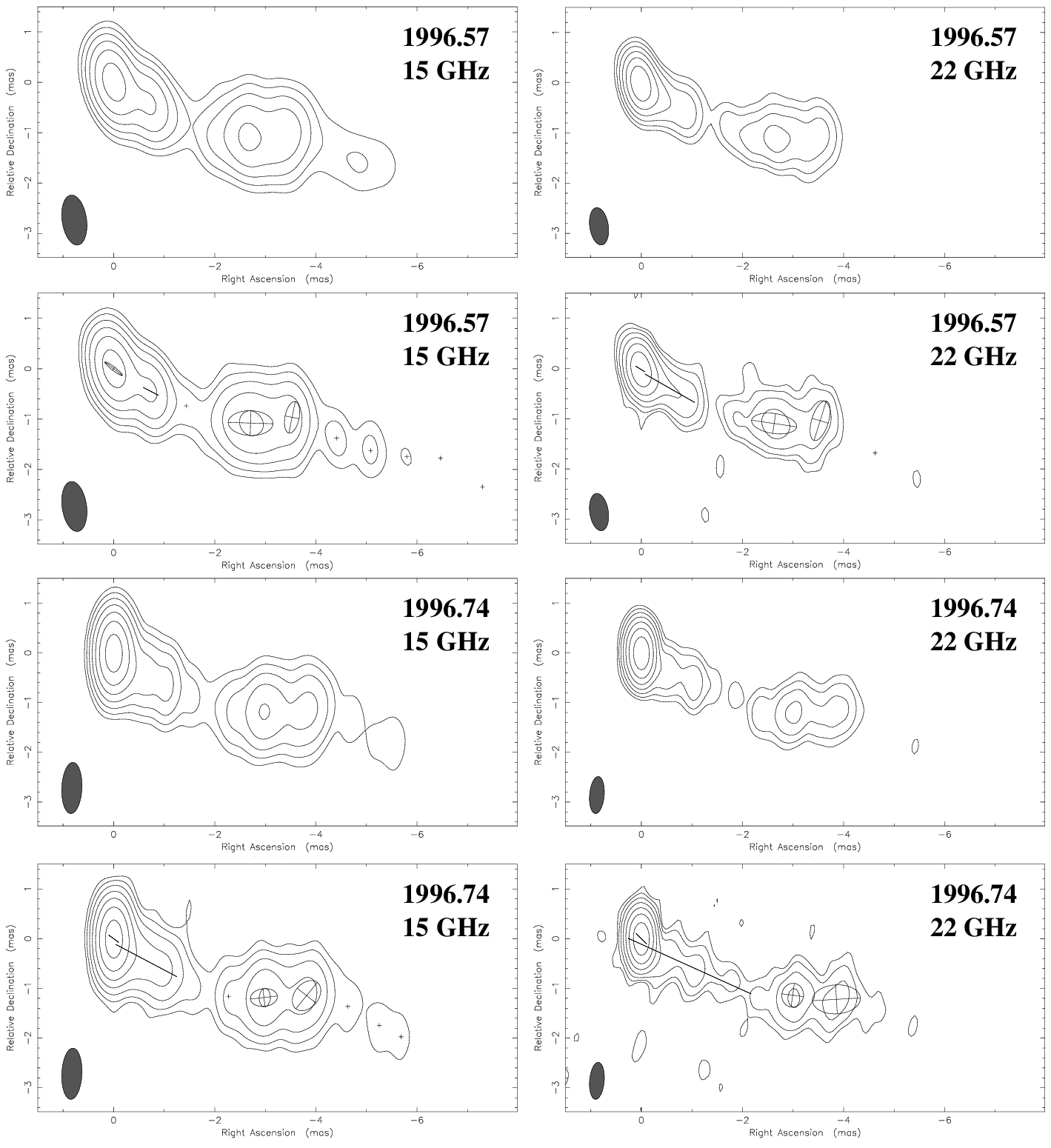}
\caption{\label{f:3C120-model}
Two epochs of 3C\,120 at two frequencies.  Images created by cleaning the 
data (first image of each pair) are directly compared to the images 
produced by restoring the model-fit components to the transform 
of the model-fit residuals (second image of each pair). Note the
essential similarity of the two sets of images, demonstrating the adequacy
of the modeling procedure. 
}
\end{figure*}

Figure \ref{f:3C120-model} displays images at two epochs of 3C\,120 at both of our
frequencies. Images created by CLEAN are compared to those produced by adding the
model-fit components to the transform of the model-fit residuals. The first thing
to note is that, even with these difficulties, the two sets of images (traditional
CLEAN and restored Gaussian components) are remarkably similar. This demonstrates
that the model-fit components are, even in this worst case example, an adequate
description of the data.  The key question is then, to what extent (if any) can the
components that we have fit be used to trace motions in the jet? Using the
criteria stated above, we find that there are three kinds of components in our fit
to 3C\,120.

First consider the two components lying in the region between $2$ and 
$4$~mas from the core, which we have labeled U1A (K1A) and U1B (K1B) in
figure \ref{f:3C120I}. They represent easily identifiable discrete structures, and
vary smoothly in position, size, flux, and polarization from epoch to epoch, and
between our two observing frequencies. Neither is very close to another bright
feature. Thus we use these two components to trace motion in the jet of
3C\,120.

Next consider the single component that we fit to the region within $2$~mas from
the core, which at every epoch is a narrow Gaussian elongated in the direction of
the jet.  While such a component does represent quite well the flux in this region
at each epoch, its parameters are not well determined, and it does not vary
smoothly in size, flux, position, or polarization from epoch to epoch, or between
frequencies. Therefore we do not attempt to extract any kinematic information from
this component.   

Finally, there are the several components that we fit beyond $4$~mas from the
core. They represent the flux distribution in this part of the jet quite well, but
we were unable to extract any kinematic information from them because
(i) we cannot reliably identify them between our two frequencies and (ii)
there are so many components in one small region that we cannot match them up from
epoch to epoch.

Hence there are three types of components that we fit to our sources. The first
are well defined, compact, isolated from each other, and clearly identifiable
across epoch and frequency. The second are also fit across epochs and frequencies,
but do not represent compact structure in the jet. Such components may be marked
on our images and discussed in \S{\ref{s:results}}, but we do not present formal
proper motions for them. The third kind are transitory (present in a few epochs
and often at only one frequency), or they are erratic in position, size, flux, and/or
polarization (i.e., they ``flicker''), or they are simply crowded too close
together. While such components surely represent real flux in the jet, we
generally do not label them in the images presented in \S{\ref{s:results}}, and
they are not included in our (very conservative) analysis of jet proper motions.

\subsection{Computing Velocities and Accelerations}
\label{s:compute}

The proper motions presented in \S{\ref{s:results}} were calculated using 
the angular component positions in $x$ and $y$ relative to 
the core component, D.  The jets are assumed to be one-sided, and D is 
taken to be the bright component at the end of the jet. By fitting the 
motions in $x$ and $y$ independently, we are sensitive to
non-radial velocities and accelerations.
We use a parameterized model for the motions in ($x$, $y$)
that allows us to simultaneously determine three quantities:

\begin{itemize}
\item The angular velocity of the component at our 
middle epoch, $t_{mid} = (t_{min}+t_{max})/2$. 
This mean velocity should correspond well to the velocity 
of the component if no acceleration was fit
to the data. 
\item The epoch of origin, $t_0$, assuming this mean 
velocity is constant with no acceleration.
\item Any acceleration of the component motion during our observations.
\end{itemize}

To measure these quantities we developed the following parameterization for motions
in ($x$, $y$):
\begin{equation}
x(t) = \mu_x(t-t_{x0}) + \frac{\dot{\mu}_x}{2}(t-t_{mid})^2
\end{equation}
\[ y(t) = \mu_y(t-t_{y0}) + \frac{\dot{\mu}_y}{2}(t-t_{mid})^2 \]

\noindent Thus $\ddot{x} = \dot{\mu}_x$ 
and $\ddot{y} = \dot{\mu}_y$ give the angular accelerations.  
The mean angular velocities, $\dot{x}(t_{mid}) = \mu_x$ and 
$\dot{y}(t_{mid}) = \mu_y$, are
given directly, as are the epochs of origin, 
$t_{x0} = t_{mid}-x(t_{mid})/\dot{x}(t_{mid})$
and $t_{y0} = t_{mid}-y(t_{mid})/\dot{y}(t_{mid})$.  
These models were fit to 
our position versus time data using a $\chi^2$ minimization routine 
\citep{PTVF95}.  
In this procedure, the positions measured at the two frequencies were 
treated independently, giving us $\approx 12$ data points for each component.  
Our computed proper motions are presented in table \ref{t:motions} as 
vector proper motions.  The reported epochs of origin are an average 
of $t_{x0}$ and $t_{y0}$ weighted by their uncertainties.

In reporting our measured accelerations in table \ref{t:motions}, we resolve the
angular accelerations into two components.  One component, $\dot{\mu}_\parallel$, is
along the nominal velocity direction, $\phi$, and represents changes in
speed of the component during our observations.  The second component, 
$\dot{\mu}_\perp$,
is taken along, $\phi+90^\circ$, and represents changes in direction of motion
during our observations.

The reader will note that the proper motion plots in \S{\ref{s:results}} 
contain no error bars on the positions of components.  We found that no 
simple rule for estimating positional uncertainties
applied well to all components or even to a majority of components.  Applying a
more detailed procedure, such as manually varying positional parameters for
each component while monitoring relative changes in the $\chi^2$ statistic,
the visual fit with the (u,v)-data, and the flux distribution in the residual
image, is a viable solution when a small number of components are involved.  
However, such a procedure still does not address uncertainties due to
fundamental changes in the model (e.g., changes in the number
of components fit in a given epoch), possible errors created during the
iterative self-calibration and imaging procedure, or jitter in the observed
core location.  Given these issues and the large number of data points
for each component, we decided to use the variance about the best fit to estimate
the uncertainties in the fitted parameters.
To do this, we initially set the uncertainty for each data point equal to 
unity.  A first pass through fitting our parameterized proper motion model gave a 
preliminary $\chi^2$ value.  Taking this preliminary $\chi^2$ value, we then 
uniformly re-scaled the uncertainties on the data points so that 
$\chi^2 = N_{points}-N_{parameters}$.  A second pass through our fitting 
procedure then gave us good uncertainty estimates for the velocity, 
epoch of origin, and acceleration in our models.  

As a check on these procedures, we also fit a simpler linear least squares model 
assuming equally weighted data (no uncertainties on individual points 
were necessary) and no acceleration. The velocities and epochs of origin 
(as well their estimated uncertainties) 
for most components were nearly the same as the values obtained by our more 
detailed procedure.  In a few components, the difference in component velocity and 
epochs of origin between the two procedures was larger, but was still within the 
estimated uncertainties.

An interesting statistic is the root mean square variation in position about our fit 
component trajectories.  We computed this statistic with the accelerations set 
to zero, so the RMS variation is a good upper bound estimate of component
position uncertainty {\em and} any jitter in the core position (relative to which
the component positions are determined).  We find the RMS variation in position
to be typically $\sim 0.05$ milli-arcseconds with a number of smaller values 
($0.02-0.03$ mas) and a few larger values ($0.1-0.2$ mas).

\subsection{Conventions and Assumptions}
\label{s:conv}
Throughout this paper the structural position angle is denoted by 
$\theta$, while $\phi$ is the proper motion position angle.  Angular
separation from the core is denoted by $x$, $y$, and 
$R = \sqrt{x^2+y^2}$.  Angular proper motion
is given by $\mu$ and angular acceleration by $\dot{\mu}$ (for definition
of $\dot{\mu}_\parallel$ and $\dot{\mu}_\perp$ see \S{\ref{s:compute}}). 
The components are labeled U1 (K1), U2 (K2) $\cdots$ on the 15 GHz 
(22 GHz) images, in order of emergence, with the earliest labeled U1
(K1). The core is labeled D. Sometimes a component that is picked up 
beyond U1 (K1) at a later epoch is labeled U0 (K0). In all calculations 
we assume a Friedman universe with $q_{0}=0.05$ and 
$H_{0}=100h$ km s$^{-1}$ Mpc$^{-1}$.  Wherever they are discussed 
in the text, results from the literature have been converted to
our choice of cosmology.


\begin{deluxetable}{ccccccccccc}
\tablewidth{0pc}
\tablecolumns{11}
\tabletypesize{\scriptsize}
\tablenum{3}
\tablecaption{Proper motion results.\label{t:motions}}
\tablehead{$Object$ & $Component$ & $N$ & $<R>$ & $<\theta>$ & $\mu$ 
& $\phi$ & $\beta_{app} h$ & $t_0$ & $\dot{\mu}_\parallel$ & $\dot{\mu}_\perp$ \\ 
&&& (mas) & (deg) & (mas/yr) & (deg) && (years) & (mas/yr/yr) & (mas/yr/yr) }
\startdata
3C\,120 
 & K1B/U1B & 6 & $2.8$ & $-112.6\pm0.3$ & $1.62\pm0.05$ & $-102.7\pm1.5$ & $2.5\pm0.1$ 
    & $1994.66\pm0.10$ & $-0.71\pm0.39$ & $0.22\pm0.32$ \\ 
 & K1A/U1A & 6 & $3.4$ & $-106.8\pm0.5$ & $2.22\pm0.07$ & $-103.7\pm2.7$ & $3.4\pm0.1$ 
    & $1994.90\pm0.09$ & $\underline{-1.37\pm0.51}$ & $-1.10\pm0.79$ \\ 
J0530+13 
 & K2/U2\tablenotemark{a} & 7 & $0.2$ & $82.7\pm2.9$ & $0.16\pm0.02$ & $43.4\pm7.3$ & $9.7\pm1.2$ 
    & $1995.91\pm0.21$ & $-0.06\pm0.07$ & $0.08\pm0.07$ \\ 
J0738+17 
 & K1/U1\tablenotemark{b} & 6 & $0.8$ & $79.6\pm0.7$ & $0.14\pm0.03$ & $60.2\pm10.7$ & $2.4\pm0.4$ 
    & $1992.23\pm1.16$ & $-0.39\pm0.21$ & $0.24\pm0.21$ \\ 
OJ\,287 
 & K3/U3 & 6 & $0.6$ & $-94.7\pm0.6$ & $1.01\pm0.07$ & $-90.4\pm0.9$ & $12.8\pm0.9$ 
    & $1995.93\pm0.10$ & $\underline{1.28\pm0.54}$ & $-0.21\pm0.12$ \\ 
J1224+21 
 & K3/U3 & 6 & $0.6$ & $-13.6\pm0.5$ & $0.65\pm0.03$ & $-9.0\pm0.8$ & $11.2\pm0.5$ 
    & $1995.80\pm0.07$ & $0.18\pm0.12$ & $0.03\pm0.04$ \\ 
 & K2/U2 & 6 & $1.7$ & $-2.5\pm0.5$ & $0.67\pm0.03$ & $-5.3\pm0.7$ & $11.6\pm0.5$ 
    & $1994.53\pm0.15$ & $0.09\pm0.12$ & \fbox{$0.24\pm0.03$} \\ 
 & K1/U1 & 6 & $4.7$ & $6.3\pm0.3$ & $0.66\pm0.10$ & $-0.6\pm3.8$ & $11.3\pm1.7$ 
    & $1989.78\pm2.14$ & $-0.63\pm0.40$ & $0.15\pm0.18$ \\ 
3C\,273 
 & K10/U10 & 5 & $0.4$ & $-118.1\pm1.5$ & $0.77\pm0.04$ & $-117.4\pm3.3$ & $5.3\pm0.3$ 
    & $1996.12\pm0.05$ & $0.45\pm0.40$ & $-0.05\pm0.04$ \\ 
 & K9/U9 & 6 & $1.3$ & $-121.9\pm0.7$ & $0.94\pm0.06$ & $-120.4\pm3.2$ & $6.5\pm0.4$ 
    & $1995.09\pm0.13$ & $-0.74\pm0.45$ & $0.15\pm0.39$ \\ 
 & K8/U8 & 6 & $2.0$ & $-118.6\pm0.5$ & $1.15\pm0.05$ & $-120.1\pm2.6$ & $8.0\pm0.3$ 
    & $1994.81\pm0.11$ & $-0.07\pm0.38$ & $-0.71\pm0.40$ \\ 
 & K7/U7 & 6 & $2.7$ & $-114.8\pm0.6$ & $1.06\pm0.08$ & $-113.6\pm4.3$ & $7.4\pm0.6$ 
    & $1993.99\pm0.30$ & $\underline{1.30\pm0.64}$ & $-0.60\pm0.60$ \\ 
 & K4/U4 & 6 & $5.1$ & $-111.3\pm0.2$ & $0.99\pm0.05$ & $-120.6\pm3.2$ & $6.9\pm0.3$ 
    & $1991.72\pm0.34$ & $-0.13\pm0.36$ & $-0.54\pm0.42$ \\ 
3C\,279 
 & K4/U4\tablenotemark{d} & 5 & $0.2$ & $-125.1\pm2.2$ & $0.17\pm0.01$ & $-142.4\pm3.4$ & $3.6\pm0.2$ 
    & $1996.09\pm0.10$ & $0.07\pm0.06$ & $0.04\pm0.05$ \\ 
 & K1/U1 & 7 & $3.1$ & $-113.8\pm0.1$ & $0.25\pm0.01$ & $-124.4\pm1.5$ & $5.1\pm0.1$ 
    & $1985.61\pm0.43$ & \fbox{$-0.06\pm0.02$} & $-0.03\pm0.02$ \\ 
J1512$-$09 
 & K2/U2 & 5 & $0.2$ & $-32.4\pm3.0$ & $0.19\pm0.06$ & $-24.5\pm14.6$ & $2.8\pm0.9$ 
    & $1995.49\pm0.52$ & $0.36\pm0.61$ & $0.07\pm0.46$ \\ 
 & K1/U1 & 6 & $1.5$ & $-28.6\pm0.3$ & $0.96\pm0.03$ & $-28.2\pm1.7$ & $14.0\pm0.4$ 
    & $1994.93\pm0.07$ & $0.30\pm0.23$ & $0.36\pm0.22$ \\ 
J1751+09 
 & K3/U3\tablenotemark{e} & 6 & $0.3$ & $31.5\pm2.3$ & $0.45\pm0.06$ & $28.2\pm5.2$ & $5.9\pm0.8$ 
    & $1995.89\pm0.12$ & $-0.26\pm0.48$ & $0.12\pm0.33$ \\ 
J1927+73 
 & K3/U3 & 6 & $0.7$ & $150.8\pm0.4$ & $0.06\pm0.02$ & $142.8\pm17.2$ & $0.8\pm0.3$ 
    & $1986.37\pm4.30$ & $0.26\pm0.16$ & $0.01\pm0.14$ \\ 
 & K2/U2 & 6 & $1.8$ & $157.3\pm0.2$ & $0.22\pm0.03$ & $164.4\pm3.1$ & $2.8\pm0.4$ 
    & $1987.15\pm1.36$ & $-0.02\pm0.22$ & \fbox{$0.30\pm0.09$} \\ 
 & K1/U1 & 6 & $2.1$ & $172.9\pm0.1$ & $0.25\pm0.02$ & $168.9\pm3.8$ & $3.1\pm0.3$ 
    & $1989.06\pm1.12$ & $\underline{0.40\pm0.18}$ & $-0.10\pm0.13$ \\ 
J2005+77 
 & U3\tablenotemark{f} & 6 & $0.2$ & $-84.3\pm4.3$ & $0.04\pm0.03$ & $-58.9\pm33.0$ & $0.5\pm0.4$ 
    & \nodata & \nodata & \nodata \\ 
 & U2\tablenotemark{f} & 6 & $0.5$ & $-93.4\pm0.7$ & $0.02\pm0.02$ & $137.5\pm62.0$ & $0.2\pm0.2$ 
    & \nodata & \nodata & \nodata \\ 
 & U1\tablenotemark{f} & 6 & $1.7$ & $-93.3\pm0.5$ & $0.20\pm0.04$\tablenotemark{g} 
    & $172.2\pm13.9$\tablenotemark{g} & $2.8\pm0.6$\tablenotemark{g} 
    & \nodata & \nodata & \nodata \\ 
\enddata
\tablecomments{The number of epochs are given by $N$. Mean-epoch values for angular radius, $<R>$, 
and structural position angle, $<\theta>$, are provided for comparison. Proper motion and 
structural position angles are measured to be positive, counter-clockwise from north. 
See \S{\ref{s:conv}} for a general discussion of our conventions and \S{\ref{s:compute}} 
for a detailed discussion of our proper motion fitting procedure.}
\tablenotetext{a}{See section \ref{s:j0530} for discussion of possible rapid motions of 
multiple components here.}
\tablenotetext{b}{After epoch 1996.23, only U-band observations contribute to computed motion.}
\tablenotetext{c}{Component un-resolved from core at U-band until 1997.94.}
\tablenotetext{d}{The first two epochs are excluded from the fit. This component has a 
strong acceleration if these epochs are considered, see \S{\ref{s:3c279}}.}
\tablenotetext{e}{Component only observed at U-band.  A variable core may influence computed motion.}
\tablenotetext{f}{Only U-band observations used to compute motion.}
\tablenotetext{g}{Apparent motion is due to a centroid shift to the south.}
\end{deluxetable} 

\section{Results}
\label{s:results}

In this section we present our observed proper motions for the well-defined
components in each source.  Complete model fitting results as well as images of 
all sources at both frequencies and all epochs will be presented in 
Ojha et al. (in prep). Here we present a single total intensity
image for each source, labeling the components that are discussed in the text.  
We also show a plot of angular radial distance, $R$, 
versus time for all components for which we present proper motions.
When appropriate to understanding a particular component's motion, we also 
present plots of the angular position ($x$, $y$) over time.  
Table \ref{t:motions} contains a summary of the computed proper motions for 
each source in our sample (see \S{\ref{s:compute}}).  
On all of our plots (($R$, $t$) and ($x$, $y$))
we overlay a projection of these fitted proper motions.  
The trajectories may be non-radial and include accelerations; 
therefore, the projections of these motions will not, in general, be 
straight lines.\footnote{Large apparent non-radial motions and accelerations, 
revealed by curvature in the ($R$, $t$) and ($x$, $y$) plots, may not be 
significant in the light of their uncertainties, given in table \ref{t:motions}.}

\subsection{3C\,120 (J0433$+$05)}
3C\,120 is a nearby example of extra-galactic superluminal motion  
\citep{SCL79}. At a redshift of just $0.033$, an observed proper 
motion of $1$ mas/yr corresponds to a speed of $1.5h^{-1}$c. Recent 
results from \citet{GMA98} at 22 and 43 GHz show up to 10 superluminal 
components with angular velocities between $1.5$ mas/yr ($2.3h^{-1}$c) 
and $3.6$ mas/yr ($5.5h^{-1}$c).  

\begin{figure*}[hbt]
\epsscale{0.4}
\plotone{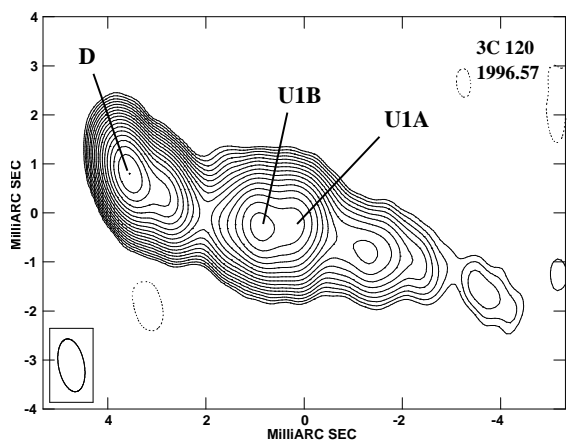}
\caption{\label{f:3C120I}
Total intensity image of 3C\,120 at 15 GHz, epoch 1996.57. Components discussed in 
the text are marked on the image.  Contours begin at 2 mJy/beam and increase 
in $\sqrt{2}$ steps.
}
\end{figure*}

\begin{figure*}[hbt]
\epsscale{0.5}
\begin{center}
\epsfig{file=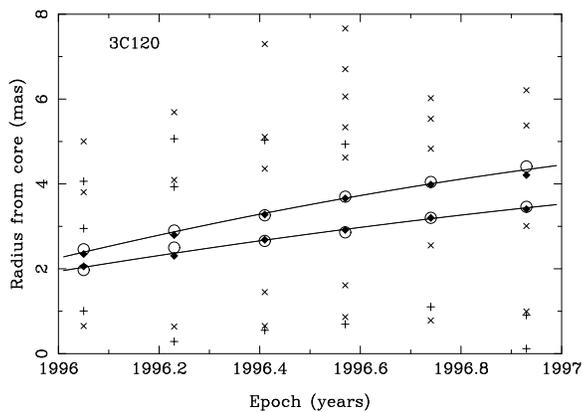,angle=-90,width=3in}
\end{center}
\caption{\label{f:3C120pl}
Radial position of model components versus time for 3C\,120. 
Components U1A (K1A) and U1B 
(K1B) are plotted with their fitted motion lines. Components are marked with 
a ``$\blacklozenge$'' at $15$ GHz and a ``$\bigcirc$'' at $22$ GHz.  Components 
that we do not follow well enough to present proper motions for are included 
on the plots marked with a ``$\times$'' at $15$ GHz and a ``$+$'' at $22$ GHz.
} 
\end{figure*}

While, in any given epoch, we observe a large number of components 
(regions of enhanced brightness) in the jet, we found it very 
difficult to identify many of these 
components across epochs and (to a lesser extent) across frequency 
(see \S{\ref{s:model}}).  The components U1A (K1A) and U1B (K1B) 
(see figure \ref{f:3C120I}), however, are unambiguously identified in 
all epochs and at both frequencies.  Our computed proper motions for 
U1A (K1A) and U1B (K1B) appear in table \ref{t:motions}. 
Figure \ref{f:3C120pl} shows the radial position of these 
components versus time with the fitted proper motions superimposed. 

\begin{figure*}[hbt]
\epsscale{0.5}
\begin{center}
\epsfig{file=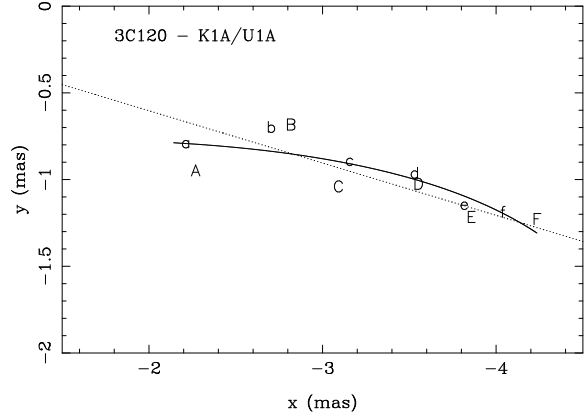,angle=-90,width=3in}
\end{center}
\caption{\label{f:3C120K1Axy}
Plot showing the $x$ and $y$ 
position from the core of component U1A (K1A) in 3C\,120. The six 
epochs are labeled A through F with the capitals referring to the higher frequency.
The solid line is the projection of the derived motion of U1A (K1A).  The dotted 
line is line for pure radial motion.  The derived motion is not significantly 
non-radial. Note the apparent deceleration in the motion of U1A (K1A).
} 
\end{figure*}

\begin{figure*}[hbt]
\begin{center}
\epsfig{file=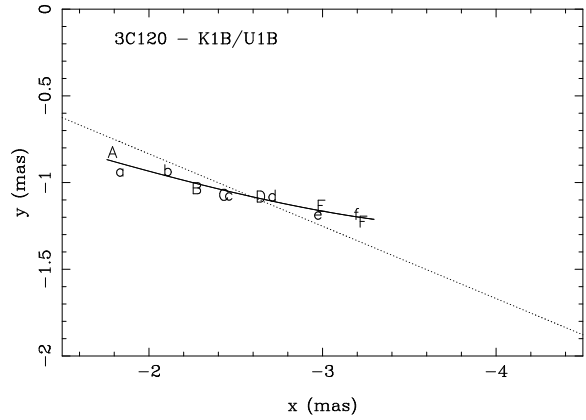,angle=-90,width=3in}
\end{center}
\caption{\label{f:3C120K1Bxy}
Plot showing the $x$ and $y$ position from the core 
of component U1B (K1B) in 3C\,120. The six 
epochs are labeled A through F with the capitals referring to the higher 
frequency. The solid line is the projection of the derived motion of U1A (K1A). 
The dotted line is line for pure radial motion.  The derived motion is 
significantly non-radial. 
} 
\end{figure*}

At $2.22\pm0.07$ and $1.62\pm0.05$ mas/yr for U1A (K1A) and U1B (K1B) 
respectively, the proper motions of these components are distinctly 
different; however, their epochs of
origin are nearly the same, straddling $1994.8$.  Figures \ref{f:3C120K1Axy}
and \ref{f:3C120K1Bxy} show the detailed ($x$, $y$) motion of these components
over time. The proper motion position angles of both components 
($\phi \simeq -103^\circ$) are nearly along the {\em local} jet direction; 
however, for component 
U1B (K1B) this proper motion position angle is $10^\circ$ from its 
structural position angle relative to the core ($-113^\circ$) and is 
distinctly non-radial. An interesting aspect of the motion of U1A (K1A) 
is that it appears to slow down over time (see figure \ref{f:3C120K1Axy}); 
we measure this deceleration to be $-1.37\pm0.51$ mas/yr/yr, 
a $2.6\sigma$ result.

\clearpage

\subsection{J0530+135 (B0528+134)}
\label{s:j0530}
This EGRET-detected quasar varies with extreme rapidity at high energies 
\citep{MDG96}. It is also well
known as a strong, flat spectrum radio source \citep{AALH85}. At
a redshift of 2.060, proper motion of $1$ mas/yr corresponds to a 
velocity $60.3h^{-1}$ times the speed of light. Global
VLBI observations at 8 and 22 GHz \citep{PRSRU96,PRKS95} found one 
component moving
at a superluminal speed of $\beta_{app}h=6.6\pm2.6$ 
($\mu = 0.11\pm0.043$ mas/yr). 
\citet{BWKQC99}, observing at 8 GHz, model up to seven jet features, 
finding superluminal motion of about $7-9h^{-1}$c ($\mu = 0.12-0.15$ mas/yr) 
for five of
them. They report that the two outermost components show higher
speeds than do the inner five. Further, they find that some of
their components move along curved trajectories that differ from
component to component. The ejection position angles of two pairs of components
are displaced by about $100^\circ$. They find that the times of ejection
coincide with the beginning of ``phases of enhanced flux-density
activity" for four components.

\begin{figure*}[hbt]
\epsscale{0.4}
\plotone{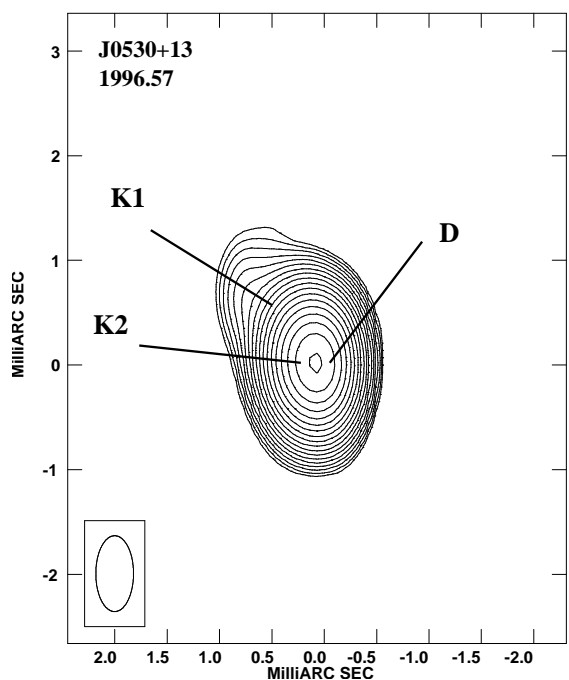}
\caption{\label{f:J0530+13I}
Total intensity image of J0530+13 at 22 GHz, epoch 1996.57.  Components 
discussed in the text
are marked in the image.  The contours begin at 15 mJy/beam and increase in 
$\sqrt{2}$ steps.
} 
\end{figure*}

Our images are the first VLBP (Very Long Baseline Polarization) images of 
this source. The total intensity structure is in agreement with 
earlier VLBI images. We see a faint jet extending up to 3 mas from the 
core.  To our data we fit a core 
and two jet components within the first milliarcsecond
of the jet. Almost all the flux is contained in the core and innermost 
component (see figure \ref{f:J0530+13I}). While this is the simplest 
model that adequately fits the total 
intensity observations, it is almost certainly not a complete description of 
the source whose complexity is more apparent in our polarized images 
(Ojha et al., in prep.).

\begin{figure*}[hbt]
\epsscale{0.5}
\begin{center}
\epsfig{file=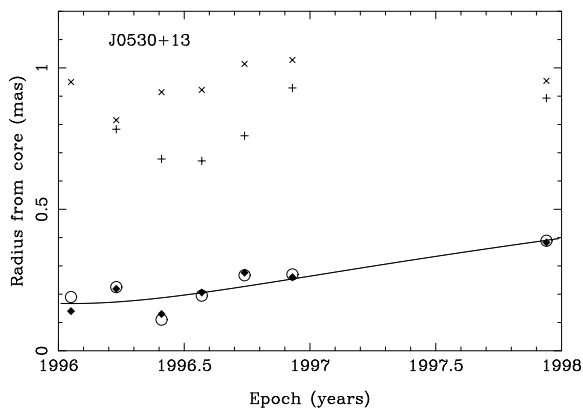,angle=-90,width=3in}
\end{center}
\caption{\label{f:J0530+13pl}
Radial position of model components versus time for J0530+13. Component K2 (U2) 
is plotted with its fitted motion line. Components are marked with a 
``$\blacklozenge$'' at $15$ GHz and a ``$\bigcirc$'' at $22$ GHz.  Components that 
we do not follow well enough to present proper motions for are included on 
the plots marked with a ``$\times$'' at $15$ GHz and a ``$+$'' at $22$ GHz.
} 
\end{figure*}

\begin{figure*}[hbt]
\begin{center}
\epsfig{file=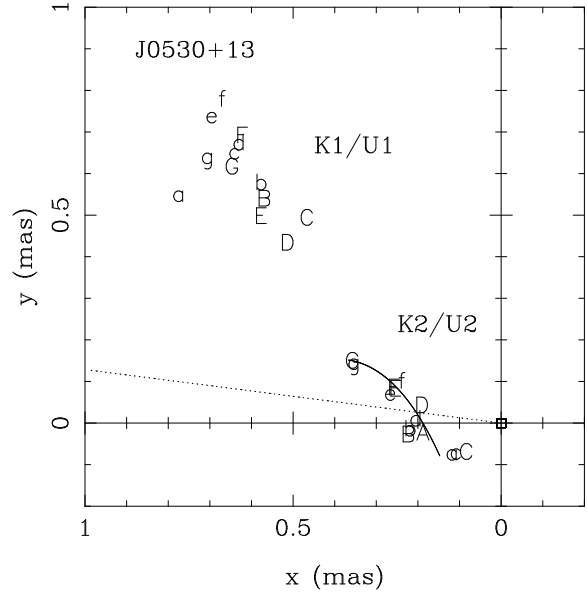,angle=-90,width=3in}
\end{center}
\caption{\label{f:J0530+13xy}
Plot showing the $x$ and $y$ 
position from the core of components K1 and K2 in J0530+13. The 
seven epochs are labeled A through G with the capitals referring to the 
higher frequency. Note that the position of K1 (upper left) varies erratically
with epoch and frequency supporting the interpretation that it is not a discrete
component. In contrast, the position of the well-defined component K2 (close to
origin) hangs together well. The non-radial nature of K2's motion stands out 
in this plot.  
} 
\end{figure*}

The outer-most component K1 (U1) has no clear proper motion
(see figures \ref{f:J0530+13pl} and \ref{f:J0530+13xy}) and is fit 
closer to the core at the higher frequency at every epoch, suggesting a 
strong spectral index gradient.
This feature is large (about 0.5 mas in size) and the differences
in position at the different frequencies are smaller than its size. Thus it
could be a single component with a large gradient in spectral index; however,
its shape and orientation vary enough that it cannot be considered a well
defined component, and we do not use it in our kinematic analysis.

The behavior of the inner component K2 (U2) lends itself to two different
interpretations. If we interpret all 7 epochs (times 2 frequencies) as 
the motion of a single component, we obtain a proper motion of $0.16\pm0.02$
mas/yr, corresponding to $\beta_{app}h = 9.7\pm1.2$ (figure \ref{f:J0530+13pl}).
The derived motion is clearly non-radial (figure \ref{f:J0530+13xy}) at a 
proper motion position angle that differs from
its mean structural position angle by at least $30^\circ$.   The proper 
motion position angle is approximately along a line from K2 (U2) toward K1 (U1).  

Examining our data, however, it is clear that our independent fits for K2 (U2) 
at 15 and 22 GHz agree very closely at every epoch from 1996.22 onward, and it seems 
unlikely that modeling error will cause the ``jittery'' variation in 
position that we see in figure \ref{f:J0530+13pl}. In addition to the 
positional variation, we see accompanying changes in the flux  of K2 (U2) 
between epochs 1996.22 and 1996.41 that suggest we may not be 
looking at the same component at the two epochs (Ojha et al., in prep.). 
Its flux rises sharply from 
$\sim$$1.5$ to $\sim$$4.0$ Jy at both frequencies, while that of the core drops 
from $\sim$$7.8$ to $\sim$$3.9$ Jy. This change in flux is not an artifact of 
our model fitting and is confirmed by changes in our polarization images. 
Taking just the epochs 1996.41 through 1996.74, we would 
fit a component moving on a well-defined trajectory with a proper motion of 
more than $20h^{-1}$c! However, at the next epoch, 1996.93, we observe K2 
(U2) falling  short of the predicted position for such a fast motion. We may 
be seeing the effects of a curving jet trajectory, multiple component ejection 
and the limits of our spatial and temporal resolution.  
We consider the simple interpretation of a single component trajectory
to be the minimum motion exhibited by the feature(s) we call K2 (U2), and
this is the motion we report in table \ref{t:motions}.

\subsection{J0738+177 (B0735+178)}
This highly optically variable BL Lac object has one of the most bent
jets on milliarcsecond scales \citep{GWRAA94}. With an absorption line redshift
of $0.424$, proper motion of 1 mas/yr corresponds to a velocity of
at least $16.9h^{-1}$c.  \citet{BZ91} reported
superluminal motion in a component of $\sim 7.9h^{-1}$c ($\sim 0.47$ mas/yr). 
\citet{GWRAA94} confirmed the motion of this component at a velocity $8.1 h^{-1}$c 
($0.48\pm0.02$ mas/yr), reported speeds 
of $5.6h^{-1}$c ($0.33\pm0.02$ mas/yr) and $4.7h^{-1}$c 
($0.28\pm0.03$ mas/yr) for two other components, and found evidence for
a stationary component with $\mu = 0.03\pm0.04$ mas/yr.
 
\begin{figure*}[hb]
\epsscale{0.5}
\plotone{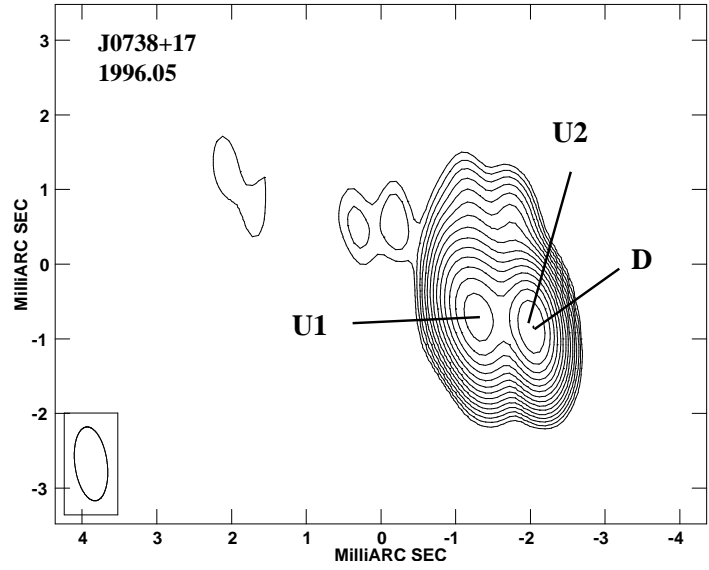}
\caption{\label{f:J0738+17I}
Total intensity image of J0738+17 at 15 GHz, epoch 1996.05.  
Components discussed in the text are
marked on the image.  Contours begin at 3 mJy/beam and increase 
in $\sqrt{2}$ steps.
} 
\end{figure*}

\begin{figure*}[hb]
\begin{center}
\epsfig{file=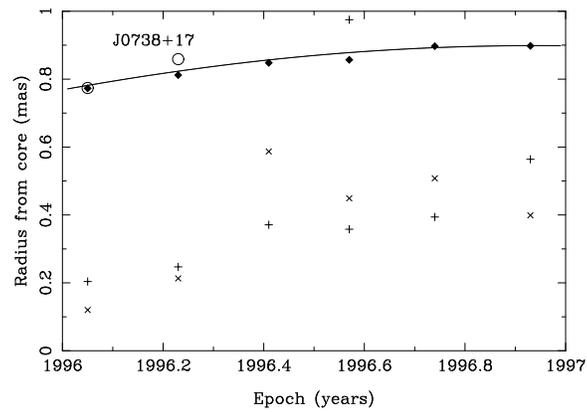,angle=-90,width=3in}
\end{center}
\caption{\label{f:J0738+17pl}
Radial position of model components versus time for J0738+17. Component U2 (K2)
is plotted with its fitted motion line. Components are marked with a 
``$\blacklozenge$'' at $15$   
GHz and a ``$\bigcirc$'' at $22$ GHz.  Components that we do not follow
well enough to present proper motions for are included on the plots
marked with a ``$\times$'' at $15$ GHz and a ``$+$'' at $22$ GHz.
} 
\end{figure*}

We see most of the flux in an East-West core-jet, plus a faint parallel ``bar" 
of emission, about 1.5 mas north of the main structure. 
This unusual morphology is also seen by \citet{KVZC98} and \citet{GMAC99}.  
Our fitting process yields one reliable jet component, U1 (K1), at $R \simeq 0.8$ 
mas and $\theta \simeq$ 80$^{\circ}$ (see figure \ref{f:J0738+17I}), that is 
consistently fit over all 
epochs at 15 GHz but only at the first two epochs at 22 GHz, probably due to 
the lack of sensitivity at the higher frequency. \citet{GMAC99} propose that 
this component marks a bend in the jet. 
However, we find no evidence of anything other than the unaccelerated, 
essentially radial motion reported in table \ref{t:motions} and 
figure \ref{f:J0738+17pl} of $\beta_{app}h = 2.4\pm0.4$ ($\mu=0.14\pm0.03$ mas/yr).

\clearpage

We also find a second component, U2 (K2), close to the core.  This component
seems to change its size, shape, flux, and position in unpredictable 
ways from epoch to epoch.  Its angular size was often much larger than 
its separation from the core and oriented along its structural position 
angle. For these reasons we do not present a proper motion for it. The 
only thing consistent about U2 (K2) is its structural position angle 
of $50-55^\circ$ which is distinctly different from that of U1 (K1).  

We were unable to fit any reproducible components to the 
bar of emission to the north, and cannot address proper motions in this 
region of the jet.

\subsection{OJ\,287 (J0854+201, B0851+202)}
\label{s:oj287}
OJ\,287 is a BL Lac object at a redshift of 0.306, where $1$ mas/yr
corresponds to an apparent speed of $12.7h^{-1}c$. 
Its VLBI structure has been studied at centimeter wavelengths by the 
Brandeis Radio Astronomy Group for about 17 years \citep{RGW87, GWR89, GC96}. 
They report a proper motion of $\sim 0.25$ mas/yr for two components. 
\citet{VCS96} report a proper motion as high as $0.4$ mas/yr for their 
component K3. \citet{TKKPBL99} report an average speed of 
about $0.5$ mas/yr for six components that they follow.
 
\begin{figure*}[hbt]
\epsscale{0.45}
\plotone{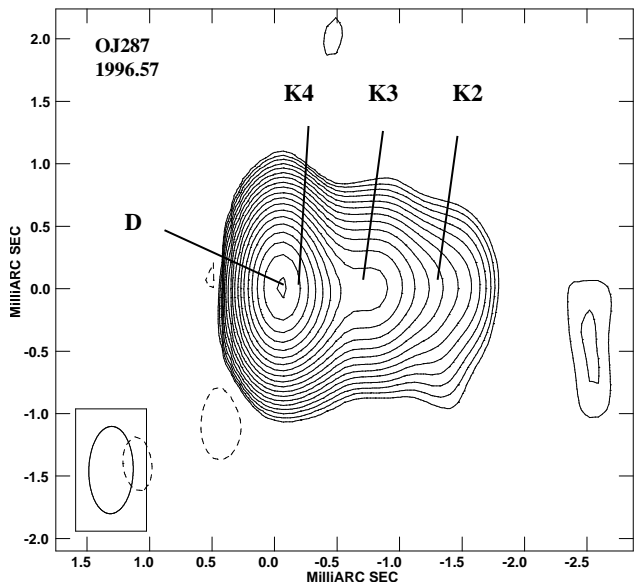}
\caption{\label{f:OJ287I}
Total intensity image of OJ\,287 at 22 GHz, epoch 1996.57.  Components 
discussed in the text are marked on the image.  Contours begin 
at 2 mJy/beam and increase in $\sqrt{2}$ steps.
} 
\end{figure*}

Our analysis of OJ\,287 provides an excellent example of the need for frequent
(no less often than bi-monthly) monitoring of such sources, as well as of the
importance of polarization images in deciphering the kinematics of
blazars. In any given epoch OJ\,287 exhibits a simple total intensity
structure, with a strong core and a short parsec scale jet extending almost
due west (figure \ref{f:OJ287I}). The structure and polarization 
of our images are consistent with the near-simultaneous 43 GHz image of 
\citet{LMG98}. The jet is well fit by two or three components
at every epoch. However, our closely-separated epochs reveal complex 
kinematics that would have been missed by less well sampled observations, 
and without the polarization images of the components, we could not have 
understood the situation with confidence. 
 
\begin{figure*}[hbt]
\begin{center}
\epsfig{file=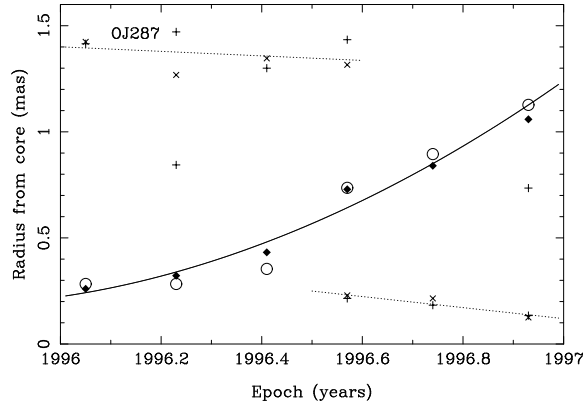,angle=-90,width=3in}
\end{center}
\caption{\label{f:OJ287pl}
Radial position of model components versus time for OJ\,287. Component K3 (U3)
is plotted with its fitted motion line (solid line). 
Components are marked with a ``$\blacklozenge$'' at $15$   
GHz and a ``$\bigcirc$'' at $22$ GHz.  Components that we do not follow
well enough to present proper motions for are included on the plots
marked with a ``$\times$'' at $15$ GHz and a ``$+$'' at $22$ GHz.
The motion of K3 (U3) is clearly affected by its proximity to 
(the as yet unresolved) component K4 in the first three epochs 
resulting in the apparent ``jump'' in its position between the third 
and fourth epochs. 
} \end{figure*}

\begin{figure*}[hbt]
\epsscale{1.0}
\plottwo{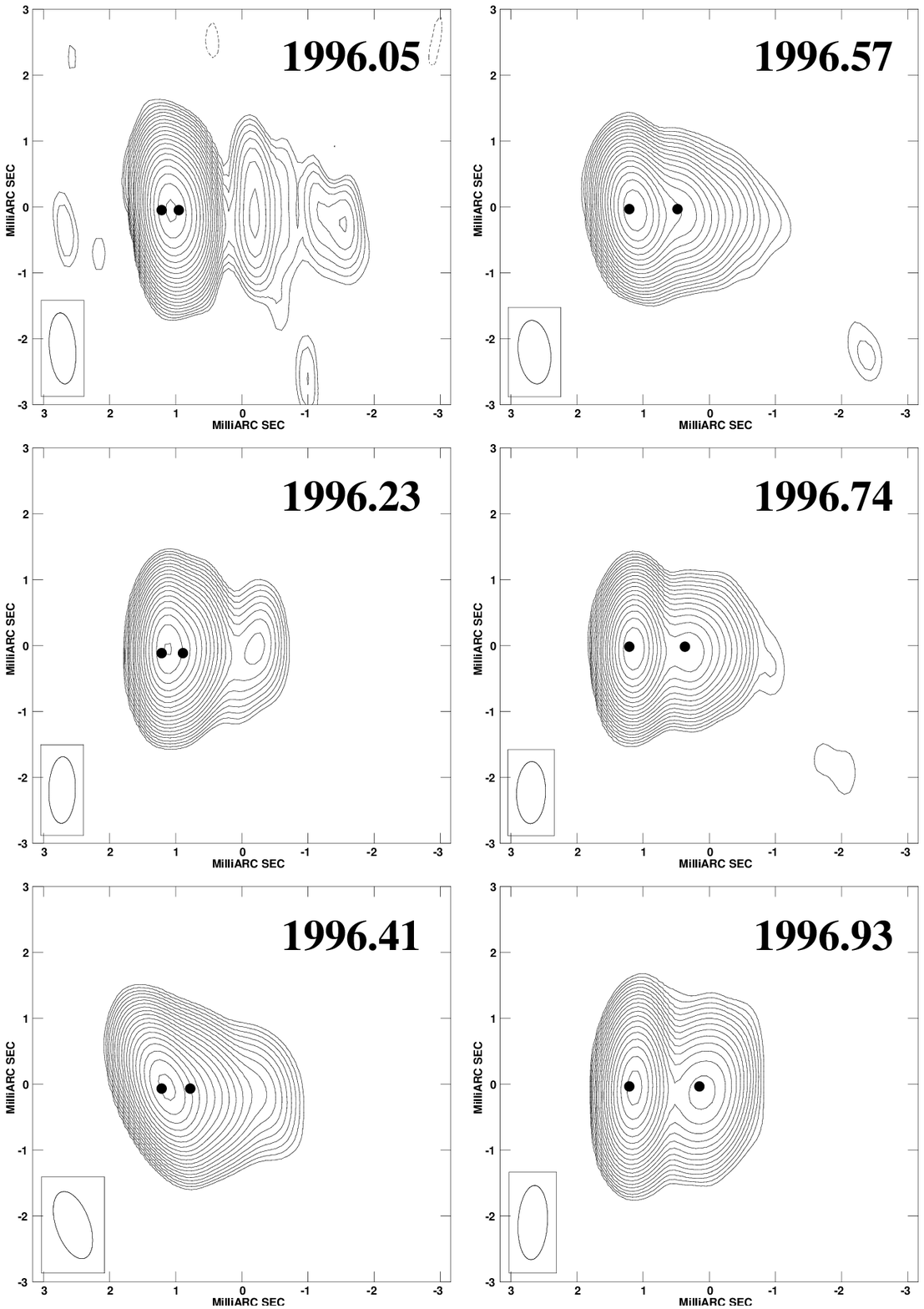}{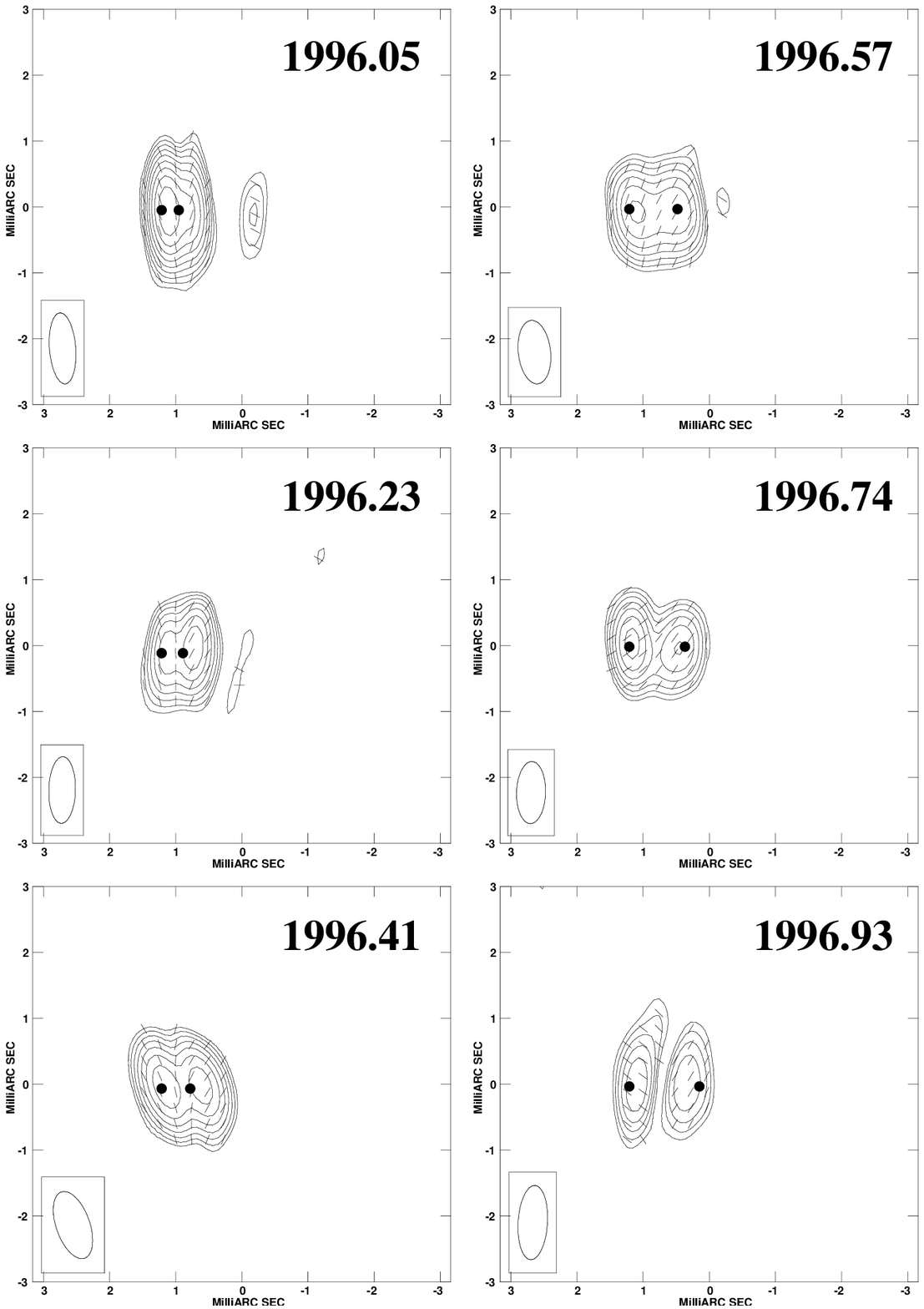}
\caption{\label{f:OJ287ip}
Total intensity (panel a) and polarization (panel b, tick marks indicate electric
vector direction) images of OJ\,287 at 15 GHz. The approximate locations of the core and 
component U3 (K3) are marked on the images.  As detailed in the text, the persistent 
polarization properties of U3 (K3) were the important clues that allowed us to 
confidently identify it across epoch.
} 
\end{figure*}

Of the three components that make up the jet, K2 (U2) and K4 (U4) show no
significant outward motion, while K3 (U3) moves radially outward with an apparent
speed of $1.01\pm0.07$ mas/yr ($\beta_{app}h = 12.8\pm0.9$) and emergence 
date $1995.93\pm0.10$ (figure \ref{f:OJ287pl}). The difficulty in discovering 
this behavior
arises from the fact that K3 (U3) passes successively through K4 (U4) and
K2 (U2) in its outward motion. For the first two epochs we do not fit K4
(U4), and for the last two epochs we do not fit K2 (U2), as during these
epochs K3 (U3) is effectively merged with K4 (U4) or K2 (U2) respectively.
This merging may bias slightly the position of K3 (U3) at these times, and 
leads to the borderline significant acceleration fit to K3 (U3). 
The key to identifying this picture is that K3 (U3) has distinctive polarization
properties, particularly its polarization position angle, that it
maintains over all the epochs (see figure \ref{f:OJ287ip}). The polarization 
position angle of K3 (U3)
is aligned neither parallel nor perpendicular to the jet axis, but at an oblique
angle to both its structural position angle and the direction of its proper motion.
The component fits agree closely between our two frequencies, both in the 
position and polarization of K3 (U3). By epoch 1997.94, K3 (U3), along with 
its peculiar mis-aligned polarization, has disappeared, which we would expect
given that its speed would place it beyond the point in the jet of OJ\,287 where 
the brightness sharply falls off at 15 and 22 GHz ($\sim 2$ mas). 

To illustrate the complex problem of discovering the motion of K3 (U3), we 
show in figure \ref{f:OJ287ip}a the 15 GHz total intensity 
images of OJ\,287 during 1996.  On each panel we have marked the approximate positions 
fit to the total intensity distributions of the core and component U3.  
We have marked these same positions in figure 
\ref{f:OJ287ip}b, which shows the polarization images for each epoch.  It is clear that
component U3 (K3) maintains the same polarization structure throughout our 
observations.  While it is true that the location of the polarization peaks do not 
precisely match those of the total intensity component, they are well within a beam 
width of one another. In the early epochs, this displacement is clearly due to ``beating'' 
with the polarization of the core component. Our component model-fits, which require 
the polarization to be fit at the location of the total intensity components and are 
less subject to ``beating'' effects, show this polarization structure to be always part 
of component U3 (K3).  
 
\subsection{J1224+212 (B1222+216, 4C\,21.35)}
\label{s:j1224}
J1224+212 is one of the most distorted of the WAT (wide-angle-tailed) quasars 
\citep{H84,SWM93}. At a redshift of 0.435, an observed proper 
motion of $1$ mas/yr corresponds to an apparent speed of $17.3h^{-1}$ 
times the speed of light. \citet{HSMB92}, report a tentative detection of
superluminal proper motion of $\beta_{app}h = 1.6 \pm 0.6$ ($0.09\pm0.04$ mas/yr) 
for a component located $\sim$$3$ mas from the core. 

\begin{figure*}[hbt]
\epsscale{0.2}
\plotone{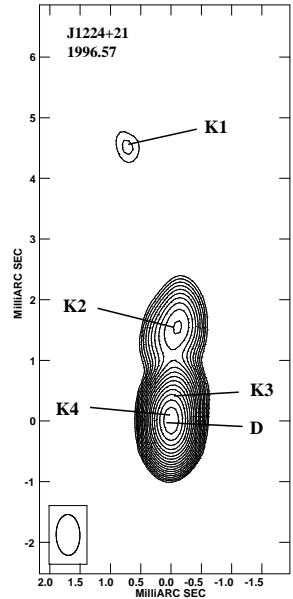}
\caption{\label{f:J1224+21I}
Total intensity image of J1224+21 at 22 GHz, epoch 1996.57.  
Components discussed in the text are
marked on the image.  Contours begin at 3 mJy/beam and increase 
in $\sqrt{2}$ steps.
} 
\end{figure*}

\begin{figure*}[hbt]
\begin{center}
\epsfig{file=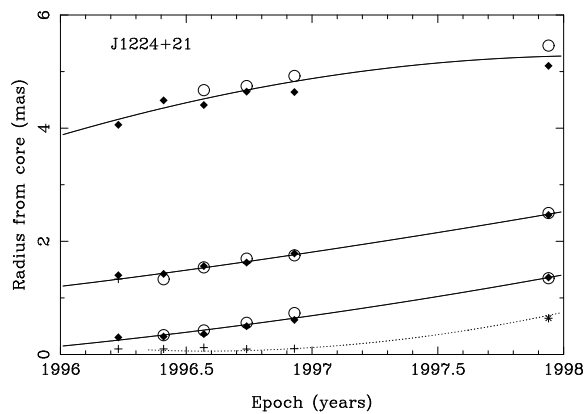,angle=-90,width=3in}
\end{center}
\caption{\label{f:J1224+21pl}
Radial position of model components versus time for J1224+21. Components K4 (U4), K3 (U3),
K2 (U2), and K1 (U1) are plotted with their fitted motion lines. 
Components are marked with a ``$\blacklozenge$'' at $15$   
GHz and a ``$\bigcirc$'' at $22$ GHz.  Components that we do not follow
well enough to present proper motions for are included on the plots
marked with a ``$\times$'' at $15$ GHz and a ``$+$'' at $22$ GHz.  
} 
\end{figure*}

\begin{figure*}[hbt]
\begin{center}
\epsfig{file=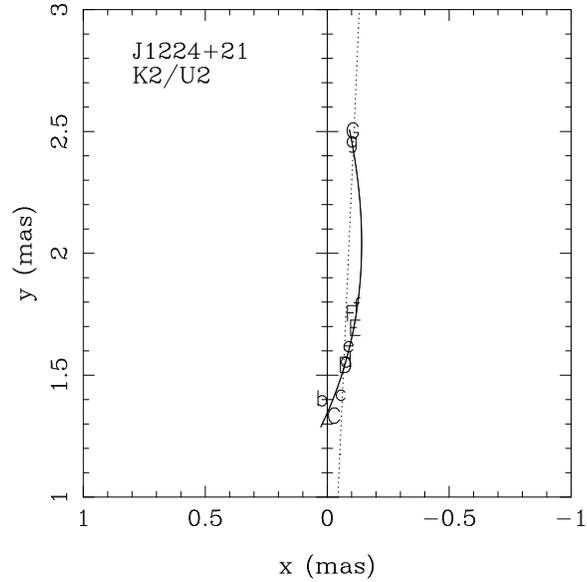,angle=-90,width=3in}
\end{center}
\caption{\label{f:J1224+21xy}
Plot showing the $x$ and $y$ position from the core of 
component K2 (U2) in J1224+21. The six epochs
are labeled B through G with the capitals referring to the higher frequency. 
The line of best fit indicates a significant perpendicular acceleration in 
the component motion.
} 
\end{figure*}

Our images show a dominant core and a north-south jet (figure 
\ref{f:J1224+21I}), in general agreement with \citet{HSMB92}. Using our two 
frequencies we are able to follow confidently the motion of three jet 
components. The proper motions of K1 (U1), K2 (U2) and K3 (U3) are highly 
superluminal with $\beta_{app}h =$ $11.3\pm1.7$, $11.6\pm0.5$ and 
$11.2\pm0.5$ respectively (figure \ref{f:J1224+21pl}).  These components 
are located at differing structural position angles increase from $-14^\circ$ to 
$+6^\circ$ with increasing radius.  The proper motion position angles of these 
components also increase from $\phi\simeq -9^\circ$ to $\phi\simeq -1^\circ$
with radius.  Interestingly, component K2 (U2) shows a significant perpendicular
acceleration $\dot{\mu}_\perp = 0.24\pm0.03$ (see figure \ref{f:J1224+21xy}) 
which is consistent with this change in position angle with radius.
Both K3 (U3) and K2 (U2) show slight, but significant, 
non-radial motions; however, the apparent position of K3 (U3) 
may be biased (particularly at U-band where U4 is not fit until 1997.94) 
by its close proximity to the core, combined with the emergence of component 
K4 (U4).  Component K4 is fit close to the core and is $\lesssim 15$\% of
the core flux; combined with its large size ($\sim 0.5$ mas) relative to its core
separation ($\sim 0.1$ mas), we did not feel that the motion of K4 could be
followed robustly during 1996.  Only in 1997.94, is K4 (U4) clearly separated
from the core.  A naive fit, including the 1996 points, is included in figure
\ref{f:J1224+21pl} as a dotted line and suggests strong acceleration
with $\mu = 0.35\pm0.01$ mas/yr and $\dot{\mu}_\parallel=0.66\pm0.09$ mas/yr/yr.
 
\subsection{3C\,273 (J1229+02)}
The famous quasar 3C\,273 is one of the best studied AGN in all wavebands.
The source has been observed with VLBI techniques for over 30 years. 
At a redshift of 0.158, a proper motion of $1.0$ mas/yr corresponds to an 
apparent proper motion of $7.0h^{-1}$ c.
\citet{ACZU96} examine all the available data in the literature including their 
own observations up to early 1991.  They find that the velocities of individual 
components do not change with time; however, the proper motions can be 
distinctly different from component to component with a range of $0.7$ to 
$1.2$ mas/yr.  

\begin{figure*}[hbt]
\epsscale{0.5}
\plotone{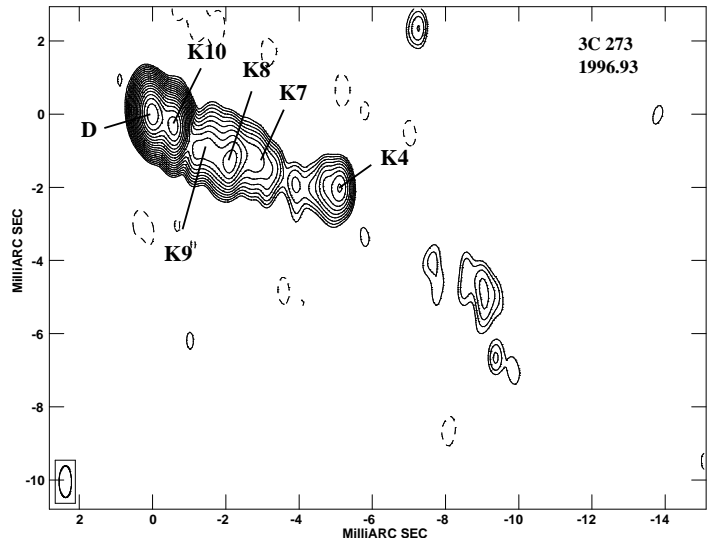}
\caption{\label{f:3C273I}
Total intensity image of 3C\,273 at 22 GHz, epoch 1996.93.  Components 
discussed in the text 
are marked on the image.  Contours begin at 20 mJy/beam and increase 
in $\sqrt{2}$ steps. 
} 
\end{figure*}

\begin{figure*}[hbt]
\begin{center}
\epsfig{file=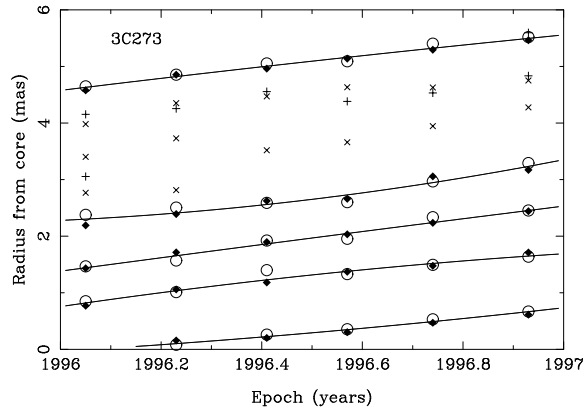,angle=-90,width=3in}
\end{center}
\caption{\label{f:3C273pl}
Radial position of model components versus time for 3C\,273. 
Components K10 (U10), K9 (U9),
K8 (U8), K7 (U7), and K4 (K4) are plotted with the projections of their 
computed motions. Components are marked with a ``$\blacklozenge$'' at $15$   
GHz and a ``$\bigcirc$'' at $22$ GHz.  Components that we do not follow
well enough to present proper motions for are included on the plots
marked with a ``$\times$'' at $15$ GHz and a ``$+$'' at $22$ GHz.
} 
\end{figure*}

We follow five components in the first 6 mas of the jet with sufficient 
confidence to compute their proper motions.  We have labeled these components 
K4, K7, K8, K9, and K10 in figure \ref{f:3C273I}, and their proper motions are 
listed in table \ref{t:motions}. Figure \ref{f:3C273pl} shows the radial 
position versus epoch for each of these components. The oldest of these 
components, K4 (U4), has an epoch 
of emergence that is later than any of the observations summarized and reported by 
\citet{ACZU96}, so no identification with the components they track is 
possible.

\clearpage

\begin{figure*}[hbt]
\begin{center}
\epsfig{file=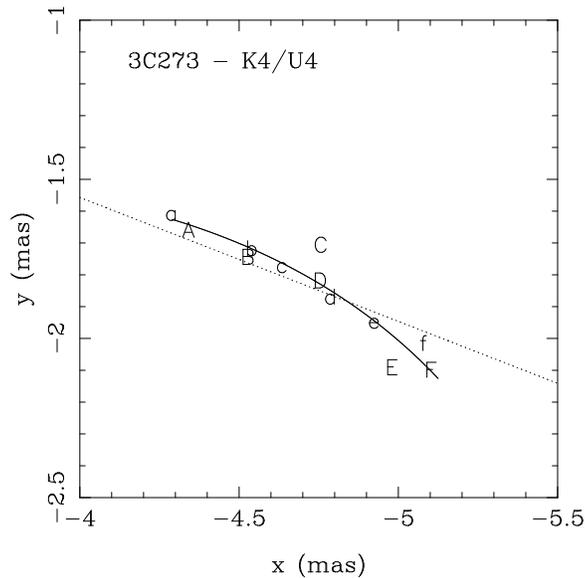,angle=-90,width=3in}
\end{center}
\caption{\label{f:3C273xy}
Plot showing the $x$ and $y$ position from the core of 
component K4 (U4) in 3C\,273. The six epochs
are labeled A through F with the capitals referring to the higher frequency. 
The line of best fit indicates borderline significant non-radial motion.
} 
\end{figure*}

As reported in table \ref{t:motions}, the proper motions of the components we 
follow range from $0.77$ to $1.15$ mas/yr ($\beta_{app}h = 5.3$ to $6.9$), 
and their structural position angles range from $-111^\circ$ to 
$-122^\circ$. A number of authors have noted a ``wiggling'' in the ridge line 
of the milli-arcsecond jet of 3C\,273 \citep{ZUCB90,K90,B91,LZD95,MJVM99}.  
While we do fit differing structural position angles for our components, it 
is interesting
that the component motions do not appear to follow this ``wiggle''.
We find that the proper motions for all components, with the
exception of K4 (U4), are consistent with radial 
motion along their structural 
position angles.  Component K4 (U4) has a fitted proper motion position angle of 
$-120.6^\circ\pm3.2^\circ$ as compared to its mean structural position angle of 
$-111.3^\circ\pm0.2^\circ$ (see figure \ref{f:3C273xy}); these values differ at 
nearly the three sigma level, suggesting non-radial motion.

\subsection{3C\,279 (J1256$-$05)}
\label{s:3c279}
Another famous quasar, 3C\,279, has also been studied for $\sim$30 years
with VLBI techniques. 
At a redshift of
$0.536$, $1$ mas/yr corresponds to an observed proper motion of
$20.6h^{-1}$ times the speed of light. 
\citet{CCGS79} measured
an expansion speed of $0.5$ mas/yr along a position angle of $-140^\circ$.
\citet{UCHZB89} follow a component they call ``C3'' with a proper
motion of $0.12\pm0.02$ mas/yr along a position angle of $-134^\circ$.  
\citet{CAUZ93} have additional observations of ``C3,'' and revise
this proper motion to $0.16\pm0.01$ mas/yr; they also observe a proper
motion for a component denoted ``C4'' of $0.15\pm0.01$ mas/yr along a
position angle of about $\sim -114^\circ$.  The same ``C4'' is apparently 
followed by \citet{UWXZM98} over a much longer period of time with
a proper motion of $\simeq 0.24$ mas/yr.  

\begin{figure*}[hbt]
\epsscale{0.35}
\plotone{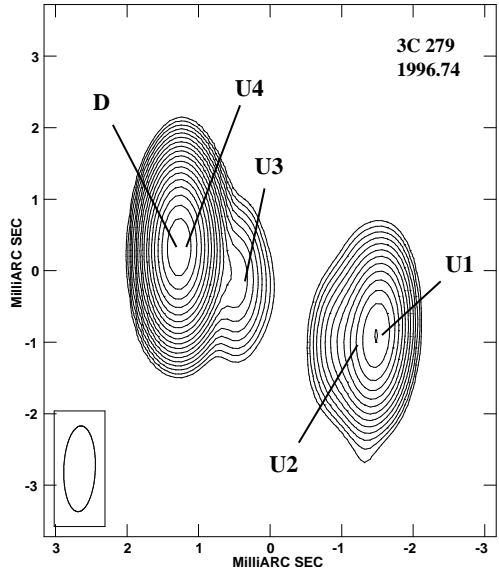}
\caption{\label{f:3C279I}
Total intensity image of 3C\,279 at 15 GHz, epoch 1996.74.  Components 
discussed in the text are
marked on the image.  Contours begin at 30 mJy/beam and increase 
in $\sqrt{2}$ steps.
} 
\end{figure*}

\begin{figure*}[hbt]
\begin{center}
\epsfig{file=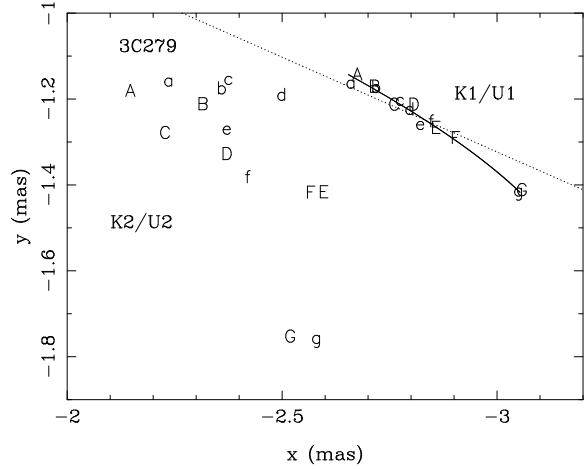,angle=-90,width=3in}
\end{center}
\caption{\label{f:3C279xy}
Plot showing the $x$ and $y$ 
position from the core of components U1 (K1) and U2 (K2) in 3C\,279. 
The seven epochs are labeled A through G, with the capitals referring to the 
higher frequency. The positions of the well-defined component U1 (K1) (upper 
right) are plotted against its projected motion which is clearly non-radial.
The positions of U2 (K2) are scattered showing how poorly this ``tail'' component
is followed.
} 
\end{figure*}

We observe only the inner $3$ mas of the jet in 3C\,279, and identify four 
components (labeled U1-U4 in figure \ref{f:3C279I}) in each of our epochs.  
Of these components, only U1 (K1) and U4 (K4) have well defined proper 
motions at both frequencies. U1 (K1) is the same component ``C4'' followed 
by \citet{UWXZM98}. Component U2 (K2) appears to be a strong but poorly 
defined ``tail'' to U1 (K1) (see figure \ref{f:3C279xy}).  Component U3 (K3) 
is not a well defined component and may represent some underlying jet emission.  

\begin{figure*}[hbt]
\begin{center}
\epsfig{file=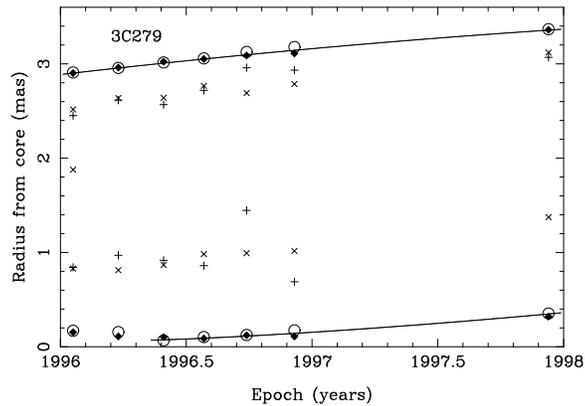,angle=-90,width=3in}
\end{center}
\caption{\label{f:3C279pl}
Radial position of model components versus time for 3C\,279. 
Components U4 (K4) and U1 (K1) 
are plotted with their fitted motion lines. 
Components are marked with a ``$\blacklozenge$'' at $15$   
GHz and a ``$\bigcirc$'' at $22$ GHz.  Components that we do not follow
well enough to present proper motions for are included on the plots
marked with a ``$\times$'' at $15$ GHz and a ``$+$'' at $22$ GHz.
} 
\end{figure*}

Our fit to the proper motion of U1 (K1) is reported in table \ref{t:motions} 
and plotted in figure \ref{f:3C279pl}.  We find a proper motion of 
$0.25\pm0.01$ mas/yr, in agreement with \citet{UWXZM98}. The proper motion of U1 (K1) 
is distinctly non-radial (see figure \ref{f:3C279xy}) with a 
proper motion position 
angle of $-124^\circ\pm2^\circ$ as compared to its mean structural position angle 
of $-114^\circ$.  This non-radial motion is paired with a slight (but significant)
deceleration of the component of $-0.06\pm0.02$ mas/yr/yr.  Both the non-radial 
motion and deceleration may result from either
an interaction of U1 (K1) with the external medium or the ``tail'' 
component U2 (K2) catching up with U1 (K1). This scenario is suggested 
by the observation that the flux of U1 (K1) rose by nearly $50\%$ from epoch 
1996.93 to epoch 1997.94 
while the flux of U2 (K2) fell off significantly (Ojha et al., in prep.).

\begin{figure*}[hbt]
\begin{center}
\epsfig{file=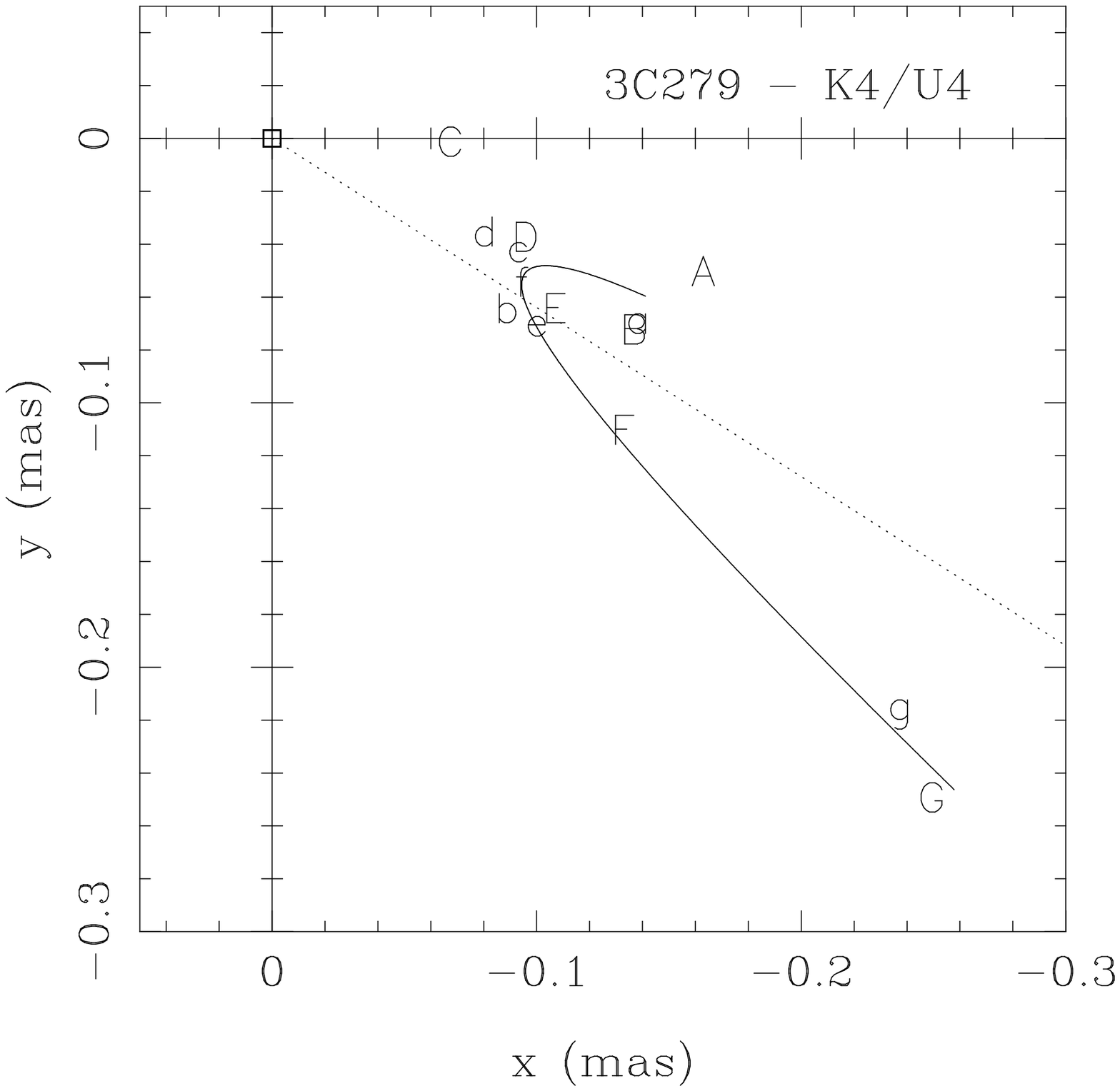,angle=-90,width=3in}
\epsfig{file=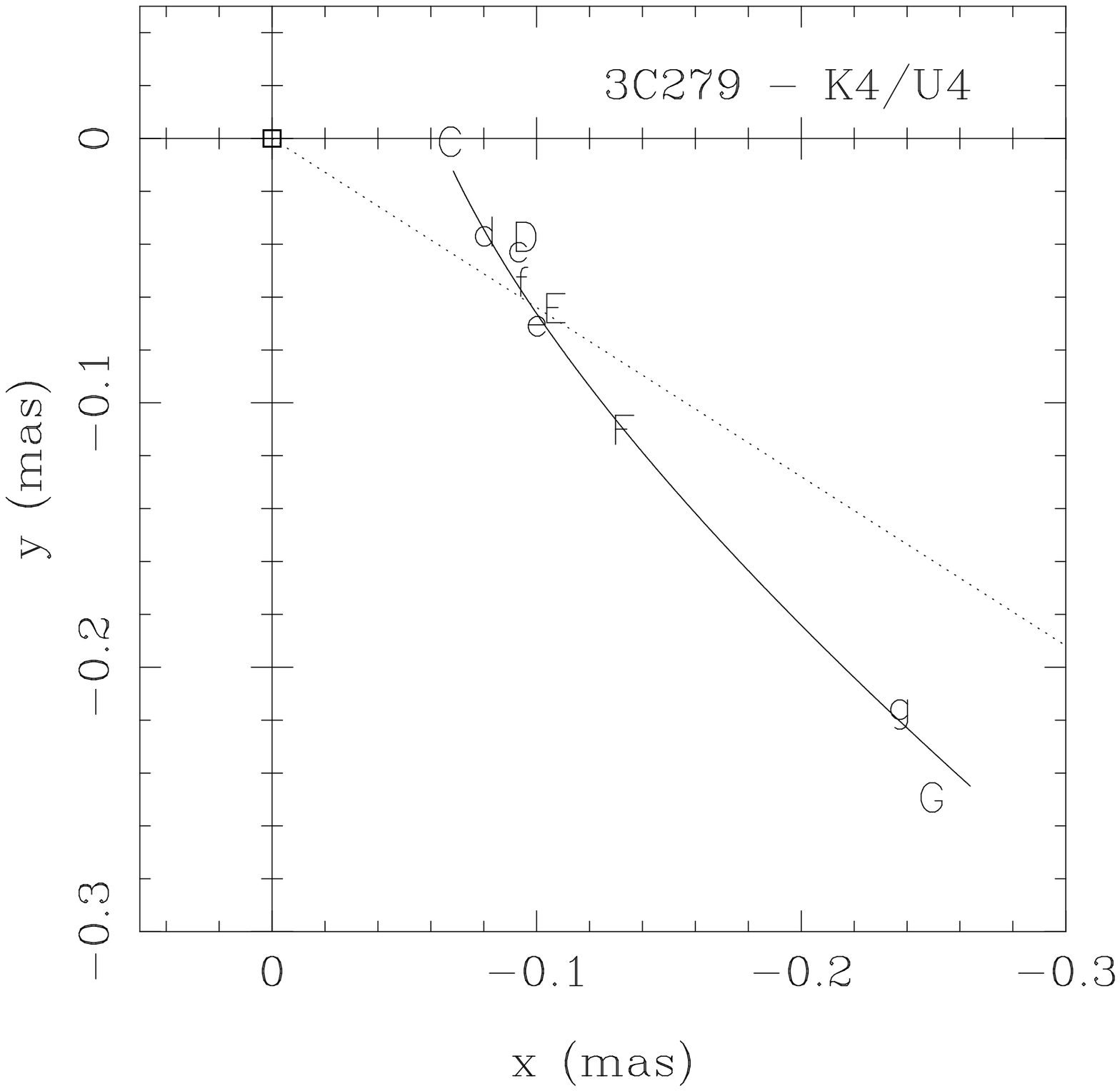,angle=-90,width=3in}
\end{center}
\caption{\label{f:3C279xy2}
Plots showing the $x$ and $y$ position from the core
of component U4 (K4) in 3C\,279. The seven epochs are labeled A through G, with the 
capitals referring to the higher frequency.  Panel (a) displays the best 
fit to all seven epochs while panel (b) excludes the first two epochs.
} 
\end{figure*}

Component U4 (K4) is a developing component very near the core
of the jet that rises sharply in flux during our observations
and has polarization properties distinct from those of the core (Ojha et al., in prep.). 
We fit what appears to be a strongly accelerated motion 
for U4 (K4), with $\dot{\mu}_\parallel = 0.24\pm0.04$ mas/yr/yr along the 
component motion, so that over the two year span of our observations, its speed 
changes by more than twice the average velocity of $0.11\pm0.01$ mas/yr.  
This component is also moving non-radially
along a position angle of $-148.2^\circ\pm5.3^\circ$ which differs from its 
structural position angle by more than $20^\circ$.  There is a slight, 
bending acceleration to the component motion of 
$\dot{\mu}_\perp = 0.08\pm0.03$ mas/yr/yr which may be related to 
this non-radial motion.  Figure \ref{f:3C279xy2}a displays this fitted motion 
on the ($x$, $y$) position plot of U4 (K4) over time.  This plot 
immediately raises some issues for our fitted motion.  It appears as though
the position of the component during the first two epochs (A and B) is inconsistent
with the motion suggested by the remaining epochs.  As stated above, component U4 (K4)
rises sharply in flux during our observations, and has its lowest flux states during
the first two epochs.  At 22 GHz, K4 is only $13$\% of the core flux in epoch A and
is only $25$\% of the core flux in epoch B. This component is large enough in 
size ($0.3-0.4$ mas FWHM) that biasing due to the proximity and strong relative 
flux of the core in these epochs is a concern.  By the later epochs, U4 (K4), is 
comparable in strength to the core, and its position is less likely to be 
seriously biased by the presence of the strong core.

In figure \ref{f:3C279xy2}b, we show the component motion without the first two
epochs contributing to the fit.  Here we find a motion $\mu = 0.17\pm0.01$ mas/yr
along a position angle of $-142.4^\circ\pm3.4^\circ$.  This motion is still 
significantly non-radial by $\approx 17^\circ$ but has no significant acceleration,
$\dot{\mu}_\parallel = 0.07\pm0.06$ mas/yr/yr. This result is more conservative
and we report it in table \ref{t:motions} and plot it in figure \ref{f:3C279pl}; however,  
it is possible that figure \ref{f:3C279xy2}a represents the correct motion for 
component U4 (K4), with the component maintaining a distance of $0.1\pm0.05$ mas 
from the core during 1996 and accelerating to $0.3$ mas by the end of 1997.
 
\subsection{J1310+323 (B1308+326, OP\,313)}
This source is a resolved blob in our VLBI images, and although
we have been able to model this source with multiple components, 
we cannot identify components over epochs or between frequencies. Hence no
information about its proper motion is presented.

\subsection{J1512$-$090 (B1510$-$089, OR\,$-$017)}
This fascinating source is discussed in detail in Wardle et al. (in 
preparation). It is one of the most violently variable examples of an OVV 
blazar \citep{MS84}. It is also classified as a radio selected quasar by 
\citet{HB93}. At a redshift of $0.360$, $1$ mas/yr corresponds to an 
observed proper motion of $14.6h^{-1}$ times the speed of light.  
\citet{FC97} have VLBI images at three frequencies that show a strong 
core containing most of the flux and a jet extending northwest.
\citet{BPFFG96} have images at three epochs with a
jet extending to the south! There is, however, an extension to the north
corresponding to the \citet{FC97} jet. 
 
\begin{figure*}[hbt]
\epsscale{0.35}
\plotone{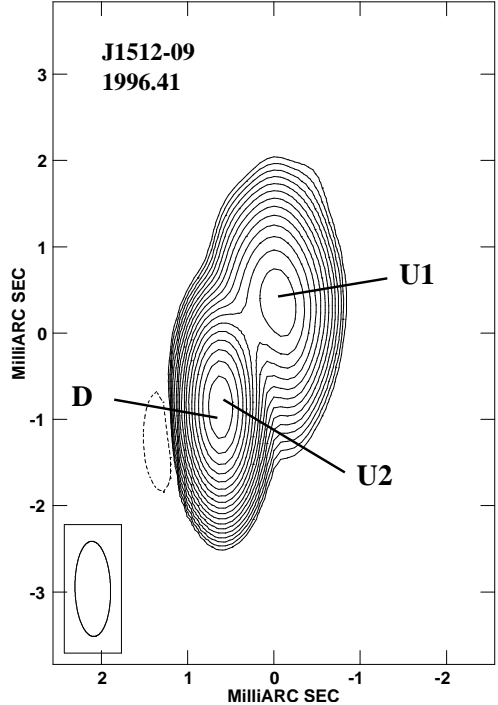}
\caption{\label{f:J1512-09I}
Total intensity image of J1512$-$09 at 15 GHz, 1996.41.  Components 
discussed in the text are
marked on the image.  Contours begin at 3 mJy/beam and increase 
in $\sqrt{2}$ steps.
} 
\end{figure*}

The total intensity structure we observe (see figure \ref{f:J1512-09I}) is similar 
to the \citet{FC97} 15 GHz image, with a strong core and a jet along 
$\theta \approx -30$ degrees.  We fit the source with a core and two jet 
components.  

\begin{figure*}[hbt]
\begin{center}
\epsfig{file=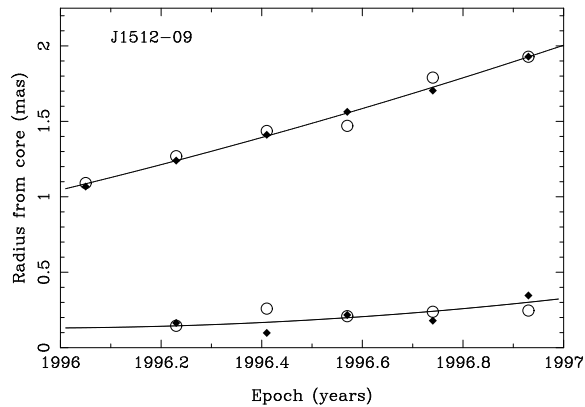,angle=-90,width=3in}
\end{center}
\caption{\label{f:J1512-09pl}
Radial position of model components versus time for J1512$-$09. 
Components U2 (K2) and U1 (K1) 
are plotted with their fitted motion lines. 
Components are marked with a ``$\blacklozenge$'' at $15$   
GHz and a ``$\bigcirc$'' at $22$ GHz.  
At $\beta_{app}h = 14.0\pm0.4$, U1 (K1) is the fastest superluminal 
component in our sample. 
} 
\end{figure*}

The proper motions for the two jet components are reported in table 
\ref{t:motions} and displayed in figure \ref{f:J1512-09pl}.  The outer jet 
component K1 (U1) is moving radially at an astonishing 
$\beta_{app}h = 14.0\pm0.4$ ($0.96\pm0.03$ mas/yr)
making it the fastest superluminal component in our sample. This 
component also appears to be expanding in angular size at superluminal speed 
(Wardle et al., in preparation). Much closer to the core, component K2 (U2) 
is also moving radially at the considerably slower speed of 
$\beta_{app}h = 2.8\pm0.9$ ($0.19\pm0.06$ mas/yr).  We detect no 
non-radial motions or accelerations in the individual motions of these
components.

\subsection{J1751+09 (B1749+096, OT\,081, 4C\,09.56)}
This compact BL Lacertae object has a redshift of $0.322$, 
so an observed proper motion of 1 mas/yr corresponds to an apparent
speed of $13.2h^{-1}$ times the speed of light. At milliarcsecond
scales over $90$\% of its flux is contained within a compact component
$\sim 0.2$ mas in size \citep{WJ80}. \citet{WCUA92} report a
strong compact source with an emission to the north. \citet{FCF96} have VLBI 
images at four frequencies also showing this northward
extension, with over 90 percent of the flux in the compact core. 
\citet{GC96} and \citet{GPC99} see a compact jet initially at 
$\theta = 25^{\circ}$ that curves to the south.  

\begin{figure*}[hbt]
\epsscale{0.3}
\plotone{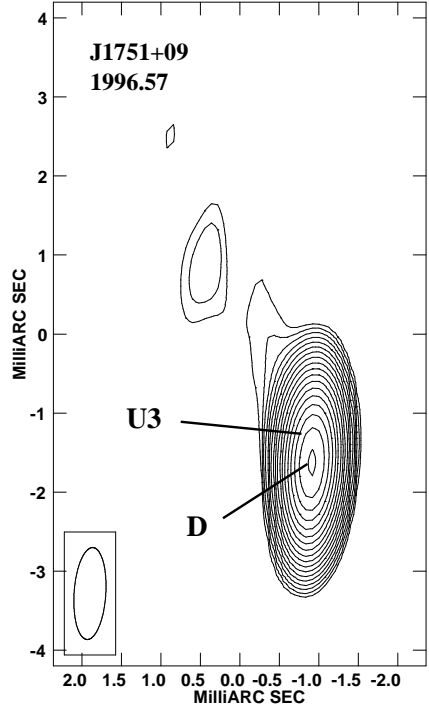}
\caption{\label{f:J1751+09I}
Total intensity image of J1751+09 at 15 GHz, epoch 1996.57.  
Components discussed in the text 
are marked on the image.  Contours begin at 2 mJy/beam and increase 
in $\sqrt{2}$ steps.
} 
\end{figure*}

\begin{figure*}[hbt]
\begin{center}
\epsfig{file=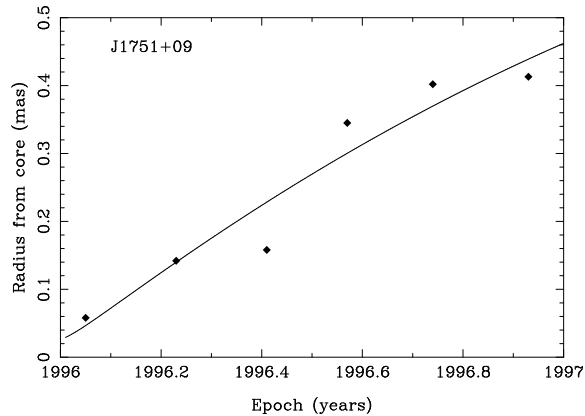,angle=-90,width=3in}
\end{center}
\caption{\label{f:J1751+09pl}
Radial position of model components versus time for J1751+09. Component U3  
is plotted with its fitted motion line. 
Components are marked with a ``$\blacklozenge$'' at $15$   
GHz.  Note the large change in position between epochs 1996.41 and 1996.57 
as U3 moves away from the influence of the strong core. 
} 
\end{figure*}

Our image (figure \ref{f:J1751+09I}) shows a very compact core containing 
over 95\% of the
flux and an extension at $\theta \simeq 30^{\circ}$. At 22 GHz we fit only a 
single Gaussian component at the core. At 15 GHz, however, 
we fit a component close to the core, labeled U3 in figure 
\ref{f:J1751+09I}, that has a proper motion of $0.45 \pm 0.06$ mas/yr 
(figure \ref{f:J1751+09pl}).  The flux of this component falls 
off very rapidly as it moves away from the core (Ojha et al., in prep.).
Because of its close proximity to the core and rapid fall-off in flux, 
the proper motion of U3 may be confused by the presence of 
the variable core.  We detect no non-radial motion or acceleration here.

\subsection{J1927+739 (B1928+738, 4C\,73.18)}
This OVV quasar is at a redshift of $0.302$ where a proper motion of
1 mas/yr corresponds to an apparent speed of $12.4h^{-1}$c. 
\citet{EWBP85,EWBJ87} and 
\citet{WSJB88}, observing at 5 GHz, find a complex jet extending 
$\sim 20$ mas to which they fit nine components of which at least five are 
superluminal, with $\beta_{app}h$ ranging from $4$ to $8$ ($0.3-0.6$ mas/yr). \citet{EWBP85} 
draw attention to the fact that the jet is not straight and the structural 
position angles 
of the components vary from $151^{\circ}$ to $175^{\circ}$ as seen from the core.
Six epochs of $22$ GHz images made by \citet{HSKR92} show three 
superluminal components, one of which may be exhibiting acceleration between 
epochs, with a velocity of about $4h^{-1}$c ($0.34$ mas/yr). They refer to work 
by Schalinski (at 5 GHz) showing an average superluminal velocity of 
$\sim 7h^{-1}$c and 
point out that taken in conjunction with their results, this indicates 
acceleration between the $\sim 2$ mas and the $6-15$ mas scales. 

\begin{figure*}[hbt]
\epsscale{0.35}
\plotone{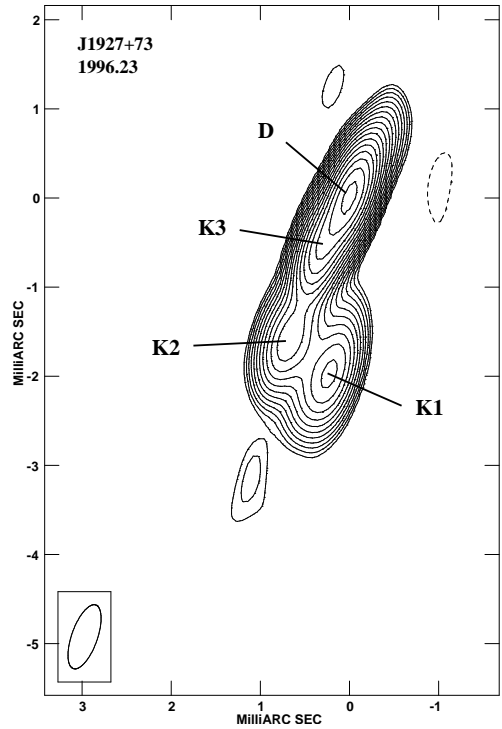}
\caption{\label{f:J1927+73I}
Total intensity image of J1927+73 at 22 GHz, epoch 1996.23.  Components discussed in the 
text are marked on the image.  Contours begin at 5 mJy/beam and increase 
in $\sqrt{2}$ steps.
} 
\end{figure*}

\begin{figure*}[hbt]
\begin{center}
\epsfig{file=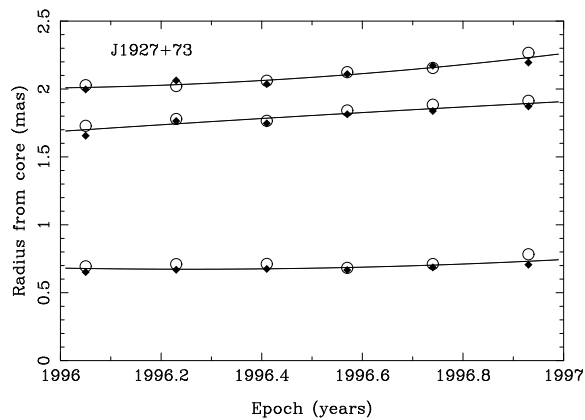,angle=-90,width=3in}
\end{center}
\caption{\label{f:J1927+73pl}
Radial position of model components versus time for J1927+73. Components K3 (U3),
K2 (U2), and K1 (U1) are plotted with their fitted motion lines. 
Components are marked with a ``$\blacklozenge$'' at $15$   
GHz and a ``$\bigcirc$'' at $22$ GHz.  
} 
\end{figure*}

\begin{figure*}[hbt]
\begin{center}
\epsfig{file=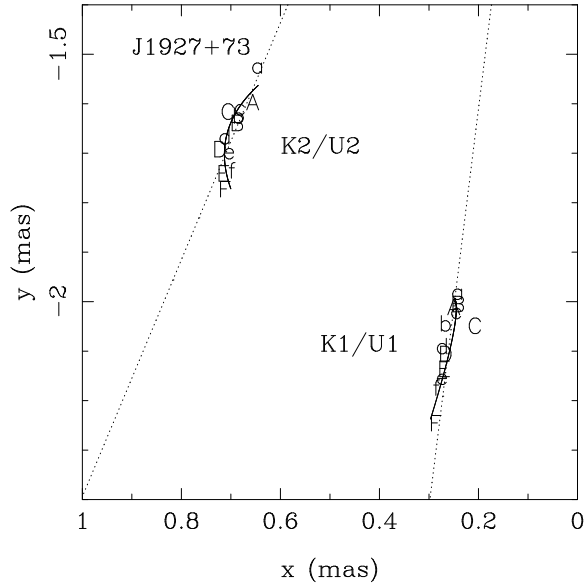,angle=-90,width=3in}
\end{center}
\caption{\label{f:J1927+73xy}
Plot showing the $x$ and $y$ 
position from the core of components K2 (U2) and K1 (U1) in J1927+73. 
The six epochs are labeled A through F, with the capitals referring to the 
higher frequency. 
} 
\end{figure*}

We fit this source with a core and three well defined jet components 
(figure \ref{f:J1927+73I}), for which proper motions are displayed in 
figure \ref{f:J1927+73pl} and reported in table \ref{t:motions}. Components K2
(U2) and K1 (U1) are located at what might be visually interpreted as two sharp 
bends in the jet; however, they are both moving predominantly radially 
(the motion of K2 (U2) has a small non-radial component which is significant 
at the $2\sigma$ level) away from the core at $\beta_{app}h = 2.8\pm0.4$ 
and $3.1\pm0.3$, respectively (figure \ref{f:J1927+73xy}). About $0.7$ mas
from the core, we fit a third component, K3 (U3), that has a 
distinctly slower radial motion of $\beta_{app}h = 0.8\pm0.3$.

\clearpage

The differing structural position angles of all three components, coupled
with their predominantly radial motions, suggests that there may be
different angles of ejection for different components in this jet.  An interesting
fact is that K2 (U2), in addition to having a slightly significant non-radial
motion, has a significant bending acceleration of $0.30\pm0.09$ mas/yr/yr,
in the right direction to bring the trajectory of K2 (U2)
parallel to that of K1 (U1).  Thus it is possible that we are seeing the effects
of collimation here.

In figure \ref{f:J1927+73xy}, it is interesting to note the location of the
components in epoch ``C'' (1996.41).  Relative to the motion suggested by the
other epochs, the ($x$, $y$) positions of both K1 (U1) and K2 (U2)
in epoch ``C'' are clearly displaced toward the core by the same amount. 
This could well be due to variation in the observed core
position, against which all the component positions are determined.
Also note that the higher frequency is systematically fit at a larger
radii for both components at most epochs.  This systematic position shift is
also apparent in figure \ref{f:J1927+73pl} for component K3 (U3).  Both pieces of
evidence suggests a core shift between $15$ and $22$ GHz of $\sim 0.04$ mas.

\subsection {J2005+778 (B2007+777)}
\label{s:j2005}
This source is unique as the only source in our sample whose
components show no discernible proper motions. The process of modeling and
understanding this source was also a reminder of the pitfalls of
modeling single frequency observations -- our 22 GHz modeling by itself
leads to an interpretation that is plausible but probably incorrect.

J2005+778 is a BL Lacertae
object at a redshift of 0.342 where a proper motion of 1 mas/yr
corresponds to an apparent speed of $14h^{-1}$c. VLBI images 
by \citet{EWBJ87} and \citet{WSJB88}
show a jet at $\theta \simeq -95^{\circ}$. They identify a
component that has a proper motion of $0.18$ mas/yr, corresponding to
$\beta_{app}h = 2.5$. \citet{GMCWR94} see a similar structure and fit
their image with six components. By identifying one of their components
with that of \citet{EWBJ87} and \citet{WSJB88}, they calculate a proper 
motion of $0.22 \pm 0.02$ mas/yr, corresponding to 
$\beta_{app}h = 3.1\pm0.8$ over $\sim 9$ years. 

\begin{figure*}[hbt]
\epsscale{0.4}
\plotone{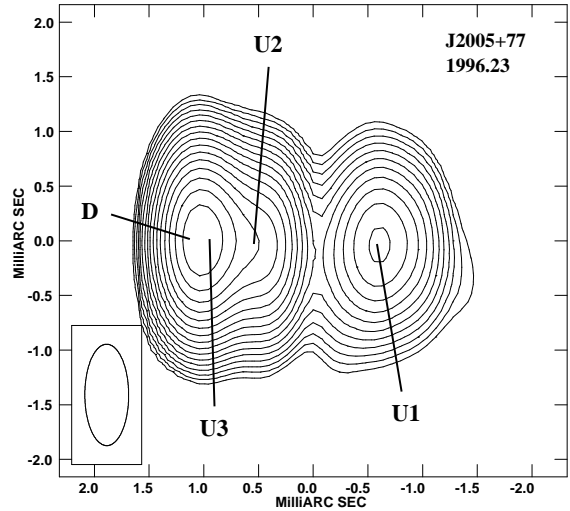}
\caption{\label{f:J2005+77I}
Total intensity image of J2005+77 at 15 GHz, epoch 1996.23.  
Components discussed in the text are
marked on the image.  Contours begin at 2 mJy/beam and increase 
in $\sqrt{2}$ steps.
} 
\end{figure*}

\begin{figure*}[hbt]
\begin{center}
\epsfig{file=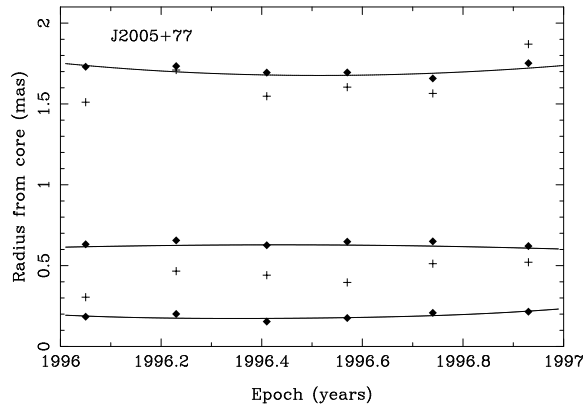,angle=-90,width=3in}
\end{center}
\caption{\label{f:J2005+77pl}
Radial position of model components versus time for J2005+77. Components U3,
U2, and U1 are plotted with their fitted motion lines.  Within
our uncertainties, these components do not propagate down the jet. 
Components are marked with a ``$\blacklozenge$'' at $15$   
GHz.  Components that we do not follow
well enough to present proper motions for are included on the plots
marked with a ``$+$'' at $22$ GHz.
} 
\end{figure*}

Our images (see figure \ref{f:J2005+77I}) show a jet extending $\sim 2.5$ 
mas  at $\theta \simeq -95^\circ$. At 22 GHz the data appear to be well 
fit with three components, the core, K1, and K2. If we confine our 
analysis to 22 GHz, we find that K1 and K2 are moving with jerky but 
distinctly superluminal apparent speeds comparable to those found by 
\citet{GMCWR94} (see figure \ref{f:J2005+77pl}).

\begin{figure*}[hbt]
\begin{center}
\epsfig{file=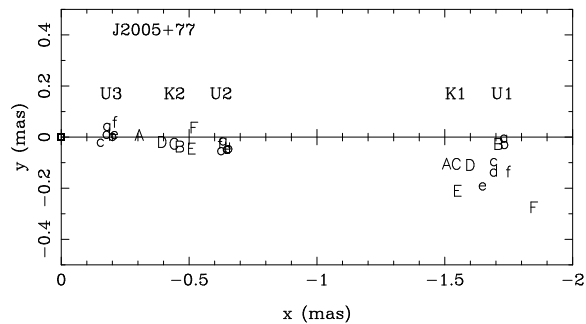,angle=-90,width=3in}
\end{center}
\caption{\label{f:J2005+77xy}
Plot showing the $x$ and $y$ 
position from the core of components U1, U2 and 
U3 in J2005+77. The six epochs are labeled A through F, with the capitals referring 
to the higher frequency. The systematic shift between components at the
two frequencies is evident. 
} 
\end{figure*}

At 15 GHz, the core region is fit with two components, D and U3. All the jet
components, U1, U2, and U3 are stationary in the radial direction (motions
are listed in table \ref{t:motions}). For
reasons we do not understand, no K3 corresponding to U3 could be fit at 22
GHz, although U3 seems to be robustly fit at 15 GHz. The radial proper motions 
observed at 22 GHz correspond very nearly to confusing the core at 22 GHz with 
U3 in the early epochs and with the 15 GHz core during the later epochs (see
figure \ref{f:J2005+77xy})! 

Thus we conclude that there is no radial motion in any component. The only 
observed proper motion is for U1 (and K1 after we remove the false radial 
motion). It appears to be moving due south at $(2.8 \pm 0.6)h^{-1}c$. This
southward motion is unlikely to be actual motion of the component. It is
probably a shift in the centroid of brightness in a large, low 
surface-brightness component. 

\section{Discussion}
\label{s:discuss}

\subsection{Component Speeds}

Out of our sample of 12 blazars, we found
proper motions in 11 sources, 10 of which are superluminal.  
Three sources (total of five components) display transverse speeds larger than 
$10 h^{-1}$c, and four other sources (seven components) display speeds larger than 
$5 h^{-1}$c. This strongly supports the AGN paradigm that requires highly 
relativistic motion near the base of a blazar jet.  The only source where
we did not detect superluminal motion is J2005+77, where none of its three 
components appear to propagate.

\begin{figure*}[hbt]
\begin{center}
\epsfig{file=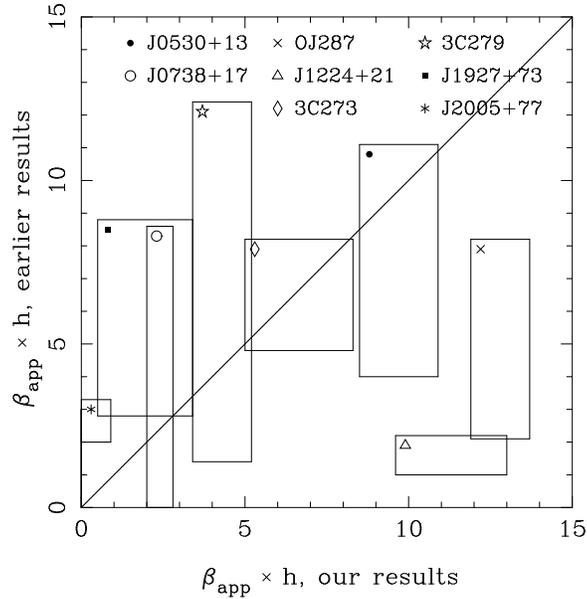,angle=-90,width=3in}
\end{center}
\caption{\label{f:oldmotions}
Diagram of shared parameter space between previously published proper 
motion results and our results on the same sources. Uncertainty estimates
are included in the range of motions, and only proper motions results with 
uncertainties $\leq 0.1$ mas/yr contribute to the plot. Previously published
results have been converted to our choice of cosmology.  Each source is 
represented by a rectangle encompassing the shared parameter space. 
} 
\end{figure*}

One goal of our program was to compare our range of observed motions to 
those published in the literature.  The range of previously published motions
is summarized in the individual source sections. In figure \ref{f:oldmotions} 
we compare this range of speeds with our range of speeds for each source.  
We find the best agreement for 3C\,273, which has been frequently observed in 
the past and to which we are able to fit and follow a large number of 
components.  It is interesting to note that we get this agreement without 
following any of the same components previously reported on in the literature.  
In general, we observe a much narrower range of motions than exist in the literature.
This may simply result from our short time baseline and the fact that we follow
a small number of components.  

There are four sources with obvious disagreement in range of motions: OJ\,287,
J1224+21, J1927+73, and J2005+77.  We find OJ\,287 to have a larger speed than previously 
observed, even though the well sampled observations of \citet{TKKPBL99} include 
our epochs. While their observations were at $8$ GHz, the transatlantic baselines in
their array gave them a resolution approaching ours at $15$ GHz with the VLBA.
They may have found only slower speeds because they did not have polarization 
observations to resolve confusion and help identify the fast moving component 
we observe.  Our observations suggest that OJ\,287 also has slower speeds 
in the jet, but we do not follow those 
components well enough to calculate robust motions.  For J1224+21, our motions
are much faster than those in the literature; however, the only previously 
published result is tentative 
and based on only two epochs.  In J1927+73, our upper limits of motions agree
with the lower limits from well sampled previous observations.  The previous 
observations are prior to 1990, and the source may simply exhibit a wide range 
of motions.  In the case of J2005+77, we also find slower proper motions 
(no motion!) than previously published. Again, variability may be an issue, as the
earlier observations were all prior to 1990; however it is also important to
note that the previous observations were at $5$ GHz which probes different 
physical scales than do ours. 

\begin{figure*}[hbt]
\begin{center}
\epsfig{file=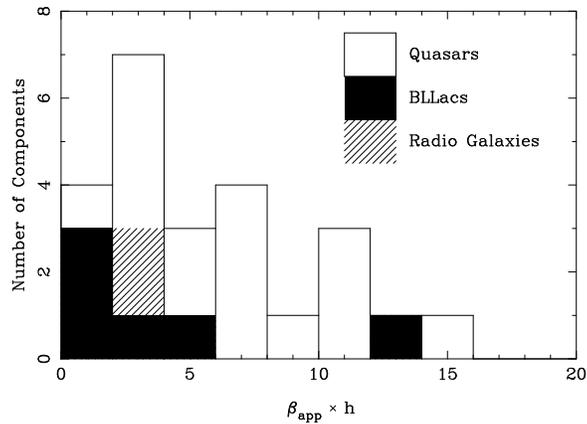,angle=-90,width=3in}
\end{center}
\caption{\label{f:hist1}
Histogram of the apparent velocity of all components in our sample of 12 blazars. 
} 
\end{figure*}

Although our source sample is small and not complete, it is interesting to 
examine the connection of apparent speed to optical identification.  
\citet{GMCWR94} find that VLBI component speeds are systematically 
slower in BL Lacs than in quasars; however, \citet{VC94} do not find 
strong evidence of such a dichotomy.  
Figure \ref{f:hist1} is a histogram of all the components for which we 
have measured proper motion.  A straight Kolmogorov-Smirnov (K-S) test shows no
significant difference between the distributions of quasar and BL Lac component 
speeds, with a probability of $0.28$ that they are drawn from the same 
distribution.  If we consider only the highest speed component in each source,
the sample sizes are simply too small ($4$ BL Lacs and $6$ quasars) for
a valid K-S test.

\subsection{Non-radial Motions}

\begin{figure*}[hbt]
\begin{center}
\epsfig{file=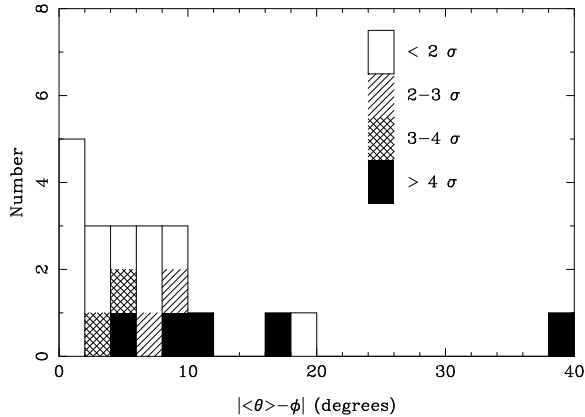,angle=-90,width=3in}
\end{center}
\caption{\label{f:nonrad_hist}
Histogram of non-radial motions in our sample of 12 blazars.  The mean structural position
angle is $<\theta>$, and the proper motion position angle is $\phi$.
} 
\end{figure*}

Non-radial motion is the motion of a component along a position angle that 
differs significantly from its structural position angle relative to the core.
Apparent non-radial motions may result from bent jet trajectories and/or 
centroid shifts due to flux or morphological changes in a component. Figure 
\ref{f:nonrad_hist} is a histogram of the non-radial motions in our observations.
Most component motion is not significantly non-radial (12 components are $< 2\sigma$).  
There are two
cases of borderline significance ($2-3\sigma$), and seven significant ($> 3\sigma$)
non-radial motions.  Four of the significant non-radial motions have 
$|<\theta>-\phi|\gtrsim 10^\circ$ 
(3C\,120 has a component at $|<\theta>-\phi| = 9.9^\circ\pm1.6^\circ$ which falls
in the bin just under $10^\circ$ in figure \ref{f:nonrad_hist}.).  Two of these
large, non-radial motions are in 3C\,279, one is in 3C\,120, and the largest 
(nearly $40^\circ$) is in J0530+13.  Examining table \ref{t:motions}, it appears
that the outer-most component in J2005+77 has a non-radial motion of nearly 
$90^\circ$; however, as explained in section \ref{s:j2005}, this motion is 
almost surely due to a centroid shift in this large, low surface brightness 
component between epochs 1996.23 and 1996.41.

Related to the issue of non-radial motion is the existence of radial motion 
of two or more jet components along distinctly different structural position angles
in a given source. The clearest case of this phenomenon is J1927+73, though 
the inner four components of 3C\,273 also appear to be moving radially at 
slightly different (but significantly so) structural position angles. 
J1224+21 also appears to have motion along slightly different position angles
in its jet; some of those motions are also slightly non-radial.  In J1224+21, the
proper motion position angles of the components increase from 
$\phi\simeq -9^\circ$ to $\phi\simeq -1^\circ$ with increasing radius.  Interestingly, 
we detected a significant, perpendicular acceleration (see below) in one of 
the components that is consistent with this change in proper motion position
angle with radius. There is also a significant acceleration which changes
the trajectory of a component in J1927+73 (see below). 

In examining the pattern of structural position angles down the jet, 
only J1224+21, 3C\,273, J1927+73, and J2005+77 have three or more jet 
components that we follow well. In J2005+77, the outer two jet 
components lie along the same position angle, while the inner-most
component differs from them by $10^\circ\pm5^\circ$.  In both J1224+21 and 
J1927+73, all three jet components lie in a steady progression of 
position angle with radius spanning $\sim 20^\circ$.  In 3C\,273, 
the outer four jet components lie in a steady progression of position 
angle with radius spanning $\sim 10^\circ$; however, this trend is 
reversed by the inner-most jet component. Taken together, these cases 
might suggest a slow oscillation in component ejection angle; however, 
with such small statistics, components may simply be located at 
random position angles within a cone of broader emission.  

\subsection{Accelerations}
\label{s:accel}

Accelerations in the motions of individual jet components have been reported
in the literature e.g., \citep{HZP96, VCS96}, and here we consider both accelerations
along a component's velocity, $\dot{\mu}_\parallel$ (speeding up or slowing down),
and perpendicular to it, $\dot{\mu}_\perp$, i.e., ``bending'' accelerations that 
change a component's trajectory.  Section \ref{s:compute} contains the details 
of our component motion fitting including acceleration.

We were unable to detect significant accelerations in most components 
(see table \ref{t:motions}).
There were four components with borderline significant ($2-3 \sigma$) accelerations
and three components with significant ($> 3\sigma$) accelerations.  
In table \ref{t:motions} we have marked the borderline and significant accelerations
by underlining or boxing, respectively.  

One difficulty we have with measuring reliable accelerations is our short total
time baseline (1 year).  It is no surprise, therefore, that two of our three
significant accelerations were found on sources for which we had an extra epoch
of observation (1997.94). We have no clear cases of individual 
components speeding up in their motions, although there are hints of such
an accelerated motion in the inner-most components of both J1224+21 and 3C\,279 
(see \S{\ref{s:j1224}} and \S{\ref{s:3c279}}).  
For both sources, we have the difficulty of confusion with the strong core which may bias
the position of the innermost jet component in some epochs.

In 3C\,279 component K1 appears to be slightly (but significantly) decelerating
in its outward motion.  Combined with the $\simeq 10^\circ$ non-radial motion
of this component, it appears that K1 (U1) is either interacting with the 
external medium or its ``tail'' component, K2 (U2), is catching up as described 
in \S{\ref{s:3c279}}.

There are two components that appear to have significant bending accelerations.
One of these is component K2 (U2) in the source J1224+21, and the acceleration is 
in the right direction to align its motion with larger scale jet structure. 
The other component with a significant bending acceleration is 
component K2 (U2) of J1927+73 which appears to have a bending acceleration directed 
toward the structural position angle of the nearby component, K1 (U1), suggesting
collimation.

\subsection{Component Speeds vs. Jet Distance}

\begin{figure*}[hbt]
\begin{center}
\epsfig{file=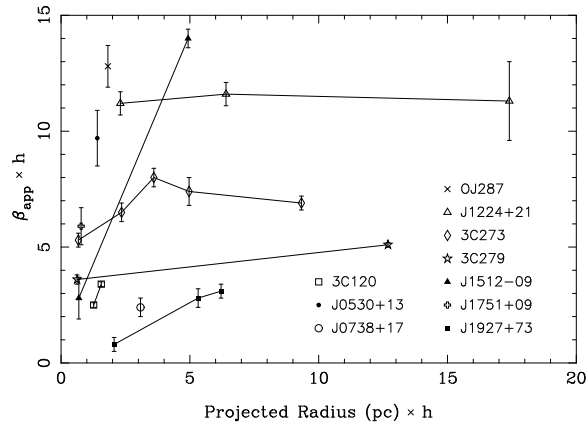,angle=-90,width=3in}
\end{center}
\caption{\label{f:mu_r}
Plot of apparent proper motion versus projected radial distance from the
core in our sample of 12 blazars.  In all six sources with multiple components, 
the innermost component is the slowest.
} 
\end{figure*}

Related to the idea of accelerations in the motions of individual components 
are different component speeds for different components down the jet.
Figure \ref{f:mu_r} is a plot of component velocity versus
projected radial distance from the core for our sample.  
Of six sources for which we report significant proper motions in more than 
one jet component, five show significantly different velocities between 
at least some components. In all five of these sources, the inner-most
component is the slowest component in the source.  This may
be evidence for systematic acceleration along the jet, perhaps
from a change in jet velocity (e.g., \citep{GM98}) or a change in
jet trajectory which may bend towards the optimum angle for 
superluminal motion.
From figure \ref{f:mu_r} any such acceleration appears to happen very 
close in, within the first few parsecs as viewed in projection.    
While there is the possibility of acceleration in the inner-most
components of J1224+21 and 3C\,279, we do not robustly observe the 
rapid acceleration of a single component near the core in any source.

\section{Conclusions}
\label{s:conclude}

We have presented proper motions for 11 highly active blazars from a 
six epoch bi-monthly monitoring program with the VLBA 
at 15 and 22 GHz.  Only one source (J2005+77) in our sample has 
no clearly observed superluminal 
components.  Of the remaining 10 sources, three (five components) have 
$\beta_{app}h > 10$ requiring $\gamma_{jet}$ of at least $10h^{-1}$ in 
these sources. Four other sources (seven components) have $\beta_{app}h > 5$.
For eight sources, we were able to compare our ranges of observed motions
to those found in the literature. Our best agreement in range was 
in 3C\,273, for which we were able to follow five superluminal components.
For four sources, we find motions well outside the range of those previously 
observed; some of our speeds or higher, others are lower.

In five of six sources for which we measure significant proper motions in 
multiple components, we see distinctly different speeds along the jet. The
innermost component is always the slowest, suggesting that acceleration
takes place along the jet.  We have no clear cases of individual 
components speeding up in their motions, although there are hints of such
an accelerated motion in the inner-most components of both J1224+21 and 3C\,279.
We do observe at least one decelerating motion and two bending 
accelerations which tend to align their motions with larger scale structure.  

We have also investigated trajectories of the moving components in our sample. 
We find most proper motion to be radial, with components in J0530+13, 3C\,120, 
and 3C\,279 being the most significant examples of non-radial motion. In at 
least two sources there are components moving radially at significantly
different structural position angles.  

Finally, we demonstrate the benefit of high frequency observations 
at closely spaced intervals to accurately measure large proper motions. There 
is clearly a strong role for multiple frequency observation and polarization 
data in elucidating complex proper motion behavior.   

\section{Acknowledgments}

This work has been supported by NASA Grants NGT-51658 and NGT5-50136
and NSF Grants AST 91-22282, AST 92-24848, AST 95-29228, and AST 98-02708.
We thank C. C. Cheung and G. Sivakoff for their help.
This research has made use of the NASA/IPAC Extragalactic Database (NED) which 
is operated by the Jet Propulsion Laboratory, California Institute of 
Technology, under contract with the National Aeronautics and Space 
Administration. This research has also made use of NASA's Astrophysics 
Data System Abstract Service.




\end{document}